\documentclass[fleqn,usenatbib]{mnras}

\usepackage{newtxtext,newtxmath}
\usepackage{graphicx}	% Including figure files
\usepackage[svgnames]{xcolor}
\usepackage{booktabs,caption}
\usepackage{threeparttable}
\usepackage{microtype}
\usepackage{gensymb}

\newcommand{\dotdeg}{\rlap{.}^\circ}

\newcommand{\dotes}{\rlap{.}^\text{s}}

\newcommand{\dotarc}{\rlap{.}''}

\DeclareUnicodeCharacter{2212}{-}

\newcommand{\review}[1]{{\color{black} #1}}

\title[SMBH of NGC~1387]{WISDOM Project -- XXVIII.\ Molecular gas measurement of the supermassive black hole mass of the galaxy NGC~1387}

\author[P.\ Dominiak et al.]{
Pandora Dominiak,$^{1}$\thanks{E-mail: pandora.dominiak@physics.ox.ac.uk}
Martin Bureau,$^{1}$\thanks{E-mail: martin.bureau@physics.ox.ac.uk}
Fu-Heng Liang,$^{1, 2, 3}$
Michele Cappellari,$^{1}$
Timothy A.\ Davis,$^{4}$
\newauthor{
Federico Lelli,$^{5}$
Ilaria Ruffa,$^{4, 5}$
Thomas G.\ Williams$^{1, 6}$
and Hengyue Zhang$^{1}$}
\\
$^{1}$Sub-department of Astrophysics, Department of Physics, University of Oxford, Denys Wilkinson Building, Keble Road, Oxford, OX1~3RH, UK\\
$^{2}$European Southern Observatory, Karl-Schwarzschild-Stra\ss e 2, Garching 85748, Germany\\
$^{3}$Astronomisches Rechen-Institut, Zentrum f\"{u}r Astronomie der Universit\"{a}t Heidelberg, M\"{o}nchhofstra\ss e 12-14, Heidelberg 69120, Germany\\
$^{4}$Cardiff Hub for Astrophysics Research \&\ Technology, School of Physics \&\ Astronomy, Cardiff University, Queens Buildings, Cardiff, CF24~3AA, UK\\
$^{5}$INAF − Arcetri Astrophysical Observatory, Largo Enrico Fermi 5, 50125, Florence, Italy\\
$^{6}$UK ALMA Regional Centre Node, Jodrell Bank Centre for Astrophysics, Department of Physics and Astronomy, The University of Manchester, Oxford Road,\\Manchester M13 9PL, UK
}

\date{Accepted XXX. Received YYY; in original form ZZZ}

\pubyear{2025}

\begin{document}
\label{firstpage}
\pagerange{\pageref{firstpage}--\pageref{lastpage}}
\maketitle

\begin{abstract}
    Supermassive black hole (SMBH) masses can be measured using molecular gas kinematics. Here we present high angular resolution ($0.12$~arcsec or $\approx11$~pc) Atacama Large Millimeter/submillimeter Array observations of the $^{12}$CO(2–1) line emission of the early-type galaxy NGC~1387. The observations reveal a face-on, regularly-rotating central molecular gas disc with a diameter of $\approx18$~arcsec ($\approx1.7$~kpc) and a central depression slightly larger than the SMBH sphere of influence. We forward model the CO data cube in a Bayesian framework with the \textsc{Kinematic Molecular Simulation} code, and use \textit{Hubble Space Telescope} data to constrain the stellar gravitational potential contribution to the molecular gas kinematics. We infer a SMBH mass of \review{$1.10^{+1.71}_{-0.95}[\text{stat},3\sigma]^{+2.45}_{-1.09}[\text{sys}]\times10^8$~M$_\odot$} and a F160W-filter stellar mass-to-light ratio of \review{$0.90^{+0.44}_{-0.35}[\text{stat}, 3\sigma]^{+0.46}_{-0.36}[\text{sys}]$~M$_\odot$/L$_{\odot,\text{F160W}}$}. This SMBH mass is consistent with the SMBH mass -- stellar velocity dispersion relation.
\end{abstract}


%%%%%%%%%%%%%%%%%%%%%%%%%%%%%%%%%%%%%%%%%%%%%%%%%%
%%%%%%%%%%%%%%%%% BODY OF PAPER %%%%%%%%%%%%%%%%%%

\section{Introduction}
\label{section:Introduction}

Despite contributing very small fractions of the total galaxy masses and having very limited spatial extents, it is now recognised that central supermassive black holes (SMBHs) have major impacts on the galaxies they lie within. Observational efforts over the last $30$ years have indeed established that SMBH mass correlates with many host galaxy properties, suggesting co-evolution \citep[e.g.][]{Gebhardt_2000, McConnell_2013, Bosch_2016}. Nonetheless, the growth of both SMBHs and galaxies and their self-regulating relationship are poorly understood \citep[e.g][]{Harrison_2018, Donofrio_2021}. 

A key issue to better understand SMBH mass -- host galaxy correlations is the accuracy of the SMBH masses. Whilst many easily-observed galactic properties can serve as indirect proxies of a SMBH's mass, an accurate measurement of the mass can only be obtained by directly probing and modelling the kinematics of the material orbiting the SMBH \citep[e.g.][]{Kormendy2013}. Traditionally, integrated stellar populations \citep[e.g.][]{Cappellari_2009, Krajnovic2009, McConnell2012}, ionised gas \citep[e.g.][]{vanderMarel_1998, Ferrarese_1996} and megamasers \citep[e.g.][]{Kuo2011, Gao2017, Pesce_2020} have been used as kinematic tracers.

Over the past decade, the unprecedented resolution and sensitivity of the Atacama Large Millimeter/submillimeter Array (ALMA) has also allowed to probe the kinematics of molecular gas within the spheres of influence (SoI) of SMBHs, thus leading to a new method of SMBH mass determination \citep{Davis_2013b}. Carbon monoxide (CO) has proven itself to be an excellent kinematic tracer, present in the nuclear regions of both early- and late-type galaxies and unattenuated by dust \citep{Ruffa_2024}.

In this work, we apply the CO dynamical modelling method to estimate the mass of the SMBH at the centre of the galaxy NGC~1387, the data for which were obtained as a part of the mm-Wave Interferometric Survey of Dark Object Masses (WISDOM) project. Thus far, the SMBH masses of $11$ galaxies have been measured as a part of the WISDOM project. The method was first tested using two early-type galaxies (ETGs) and Combined Array for Research in Millimeter Astronomy data \citep{Davis_2013b, WISDOM_I}. WISDOM work thereafter used high-resolution ALMA observations, typically of the $^{12}$CO(2–1) line, to measure the SMBH masses of a further $9$ galaxies: eight typical ETGs \citep{Davis2017, WISDOM_III, WISDOM_IV, WISDOM_V, Smith_2021, Ruffa_2023, Dominiak_2025, WISDOM_XXII, WISDOM_XXV}, one dwarf ETG \citep{Davis_2020} and one luminous infrared galaxy with an active galactic nucleus \citep{Lelli_2022}. Other groups have presented molecular-gas SMBH mass measurements using analogous methods in $13$ other ETGs \citep{Barth2016, Boizelle2019, Nagai2019, Ruffa_2019, Boizelle_2021, Cohn_2021, Kabasares_2022, Nguyen_2022, Dominiak_2024_MASSIVE, Kabasares_2024}, $2$ red nugget relic galaxies \citep{Cohn_2023, Cohn_2024} and $3$ late-type galaxies (LTGs; \citealt{Onishi_2015, Nguyen_2020, Nguyen_2021}).

This paper is structured as follows. In Section~\ref{section:NGC1387} we summarise the main properties of the target. A detailed overview of other properties of NGC~1387 is presented in \citet{Liang_2024}, a parallel paper released as a part of the WISDOM project that investigates giant molecular clouds in NGC~1387. In Section~\ref{section:ALMA Observations} we present the ALMA data and their reduction, the creation and properties of a high-resolution CO data cube and the continuum emission. We describe the dynamical modelling of the molecular gas in Section~\ref{section:Dynamical Modelling}, where we also present our results. We discuss these results in Section~\ref{section:Discussion} and summarise and conclude in Section~\ref{section:Conclusion}.

%%%%%%%%%%%%%%%%%%%%%%%%%%%%%%%%%%%%%%%%%%%%%%%%%%%%%%%%%%%%%%%%%%%%%%%%%%%%%%%%%%%%%%%%%%%%%%%%%%%%%%%%%%%
%%%%%%%%%%%%%%%%%%%%%%%%%%%%%%%%%%%%%%%%%%%%%%%%%%%%%%%%%%%%%%%%%%%%%%%%%%%%%%%%%%%%%%%%%%%%%%%%%%%%%%%%%%%

\section{NGC~1387}
\label{section:NGC1387}

NGC~1387 is a barred lenticular (SB0) galaxy located at $03^{\text{h}}36^{\text{m}}57\dotes06$, $-35\degree30\arcmin23\farcs9$ (J2000.0) within the Fornax Cluster. It is located at an angular distance of $\approx19$~arcmin ($\approx107$~kpc) to the west of NGC~1399 \citep{Napolitano_2022}, the central galaxy of the Fornax Cluster. Throughout this paper we adopt a distance of $19.3\pm0.8$~Mpc, calculated using surface brightness fluctuations \citep{Blakeslee_2009}. At this distance, $1$~arcsec corresponds to $\approx94$~pc.

\begin{figure*}
    \centering
    \includegraphics[scale=0.53, trim={0.5cm 0 0 0}]{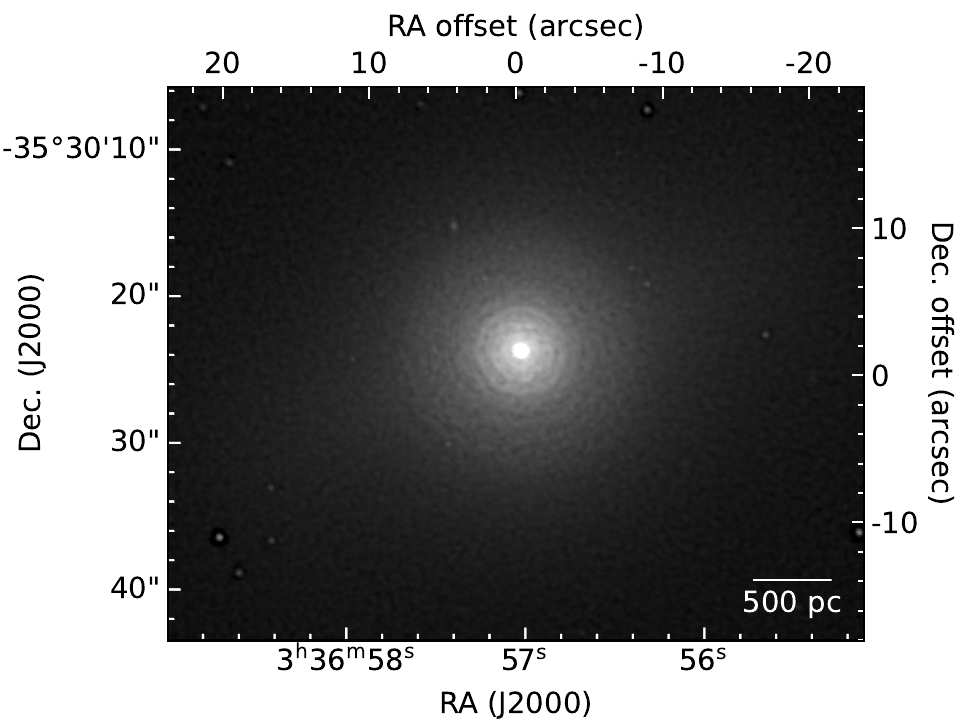}
    \includegraphics[scale=0.53, trim={0.0cm 0 0 0}]{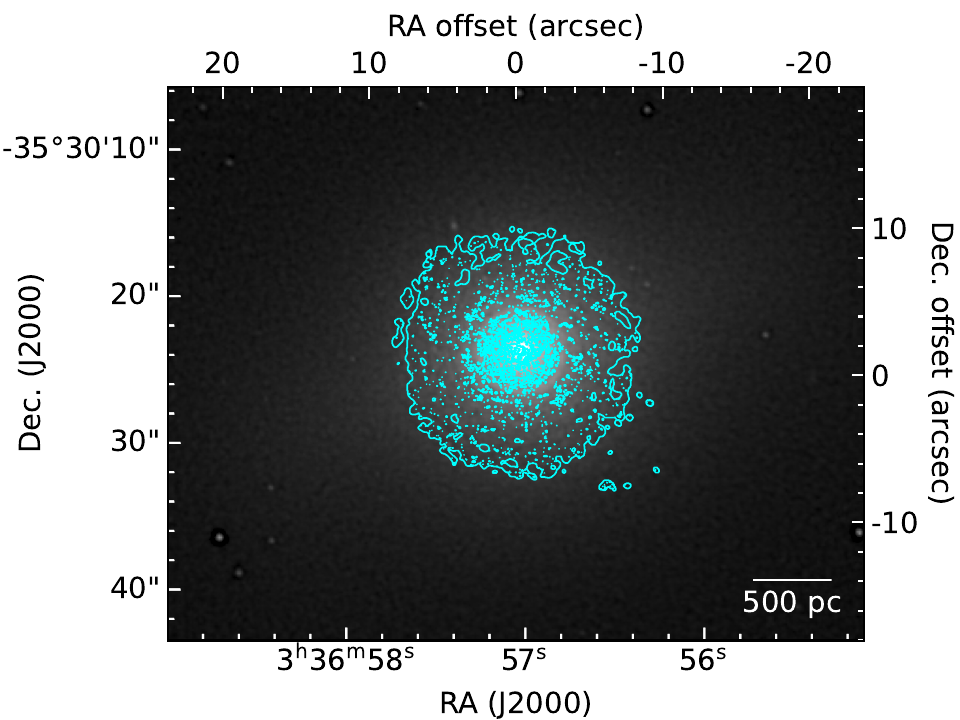}
    \caption{Unsharp-masked \textit{HST} Wide Field Camera~3 F160W-filter ($H$-band) image of NGC~1387 (left), overlaid with the $^{12}$CO(2–1) integrated intensity contours (cyan) from our high-resolution ALMA data cube (right). Note that the unsharp masking has removed the smooth outer galaxy light and instead highlights the face-on nuclear disc and bar of the galaxy.}
    \label{fig:HST}
\end{figure*}

NGC~1387 has a total stellar mass of $4.70 \times 10^{10}$~M$_\odot$, derived using an $i$-band absolute magnitude of $-21.81$ and a stellar mass-to-light ratio ($M/L$) in the $i$-band $M/L_i=1.31$~M$_\odot/\text{L}_{\odot,i}$ \citep{Iodice_2019}. \textit{Hubble Space Telescope} (\textit{HST}) optical images reveal a prominent central bulge, a weak bar and a face-on dust disc extending $\approx18$~arcsec ($\approx1.7$~kpc) in diameter along the major axis, with modest and flocculent dust (see Fig.~\ref{fig:HST}).

NGC~1387 does not have an existing SMBH mass measurement in the literature, but an estimate can be obtained using the relationship between SMBH mass and stellar velocity dispersion within one effective (i.e.\ half-light) radius ($M_\text{BH}$ -- $\sigma_\text{e}$ relation) of \cite{Bosch_2016}. Adopting 
$\sigma_\text{e}=170.2^{+12.2}_{-11.4}$~km~s$^{-1}$ \citep{Wegner_2003}, a SMBH mass of $\approx9^{+4}_{-3}\times10^7$~M$_\odot$ is expected. The radius of the sphere of influence ($R_\text{SoI}$) of a SMBH quantifies the spatial scale over which the gravitational potential of the SMBH is dominant. One way it can be estimated is by using $R_\text{SoI}\equiv GM_\text{BH}/\sigma_\text{e}^2$, where $G$ is the gravitational constant, yielding $R_\text{SoI}\approx12$~pc ($\approx0.13$~arcsec) for NGC~1387.

%%%%%%%%%%%%%%%%%%%%%%%%%%%%%%%%%%%%%%%%%%%%%%%%%%%%%%%%%%%%%%%%%%%%%%%%%%%%%%%%%%%%%%%%%%%%%%%%%%%%
%%%%%%%%%%%%%%%%%%%%%%%%%%%%%%%%%%%%%%%%%%%%%%%%%%%%%%%%%%%%%%%%%%%%%%%%%%%%%%%%%%%%%%%%%%%%%%%%%%%%

\section{ALMA Observations}
\label{section:ALMA Observations}

The $^{12}$CO(2–1) emission line of NGC~1387 was observed with ALMA in band~6 using both the $12$-m array and the $7$-m Atacama Compact Array (ACA) as a part of projects 2016.1.00437.S (PI: Davis) and 2016.2.00053.S (PI: Liu), respectively. The $12$-m array data are comprised of three tracks taken on 2016 November 4, 2016 December 21 and 2017 September 7 for a total of $20$~min on source. The baselines range from $15$~m to $7.5$~km, yielding a maximum recoverable scale (MRS) of $5.9$~arcsec ($\approx0.6$~kpc). The field of view (FoV) is $24.4$~arcsec ($\approx2.3$~kpc). The ACA data are comprised of one track taken on 2017 July 28 for a total of $17$~min on source. The baselines range from $9$ to $45$~m, yielding a MRS of $28.4$~arcsec ($\approx2.7$~kpc). The FoV is $41.8$~arcsec ($\approx3.9$~kpc). The properties of the observations are summarised in Table~\ref{tab:Observing Tracks}.

\begin{table*}
	\centering
	\caption{ALMA observations properties.}
	\label{tab:Observing Tracks}
	\begin{tabular}{lccccccl}
		\hline
		Track & Date & Array & Baseline range & On-source time & MRS & FoV & Calibration\\
            & & & & (s) & (arcsec, pc) & (arcsec, kpc)\\
            (1) & (2) & (3) & (4) & (5) & (6) & (7) & (8) \\ 
		\hline
		{\fontfamily{cmtt}\selectfont uid\_A002\_Xc44eb5\_X2e5d} & 07.09.2017 & 12-m & $41$~m -- $7.5$~km & $\phantom{1}726$ & $\phantom{0}1.1$, $\phantom{0}103$ & $24.4$, $2.3$ & Pipeline\\
  
		{\fontfamily{cmtt}\selectfont uid\_A002\_Xba26cb\_X492} & 04.11.2016 & 12-m & $19$~m -- $1.1$~km & $\phantom{1}242$ & $\phantom{0}3.1$, $\phantom{0}291$ & $24.4$, $2.3$ & Pipeline\\
  
		{\fontfamily{cmtt}\selectfont uid\_A002\_Xbbfdf3\_X1224} & 21.12.2016 & 12-m & $15$ -- $492$~m & $\phantom{1}242$ & $\phantom{0}5.9$, $\phantom{0}555$ & $24.4$, $2.3$ & Pipeline\\

        {\fontfamily{cmtt}\selectfont uid\_A002\_Xc2bb44\_X2142} & 28.07.2017 & 7-m & $9$ -- $45$~m & $1028$ & $28.4$, $2670$ & $41.8$, $3.9$ & Pipeline\\
		\hline
	\end{tabular}
    \begin{tablenotes}
           \item \textit{Notes.} Columns: (1) Track ID. (2) Observation date. (3) ALMA array. (4) Minimum and maximum baseline length. (5) Total on-source integration time. (6) Maximum recoverable scale, i.e.\ the largest angular scale that can be recovered with the given array configuration. (7) Field of view, i.e.\ the primary beam full width at half maximum. (8) Calibration method.
    \end{tablenotes}
\end{table*}

Each of the four observing tracks had a total of four spectral windows, one of which was centred on the redshifted frequency of the $^{12}$CO(2–1) line (rest frequency $\nu_\text{rest}=230.5380$~GHz). For the three $12$-m array tracks, this spectral window had a bandwidth of $1.875$~GHz ($\approx2440$~km~s$^{-1}$) subdivided into $3840$ channels of width $\approx488$~kHz ($\approx0.64$~km~s$^{-1}$). For the ACA track this spectral window had a bandwidth of $2.0$~GHz ($\approx 2600$~km~s$^{-1}$) subdivided into $2048$ channels of width $\approx977$~kHz ($\approx1.27$~km~s$^{-1}$). The remaining spectral windows for both the $12$-m array and the ACA tracks had a bandwidth of $2.0$~GHz ($\approx2600$~km~s$^{-1}$) subdivided into $128$ channels of width $\approx16$~MHz ($\approx20$~km~s$^{-1}$). These were used to map the continuum emission. 

The data were reduced using the \textsc{Common Astronomy Software Applications} (\textsc{casa}) package version 4.7.2 \citep{McMullin} and the standard ALMA pipeline. The following imaging steps were carried out using \textsc{casa} version 6.4.3.

%%%%%%%%%%%%%%%%%%%%%%%%%%%%%%%%%%%%%%%%%%%%%%%%%%%%%%%%%%%%%%%%%%%%%%%%%%%%%%%%%%%%%%%%%%%%%%%%%%%%
\subsection{Continuum emission}

\begin{figure}
    \centering
    \includegraphics[scale = 0.44]{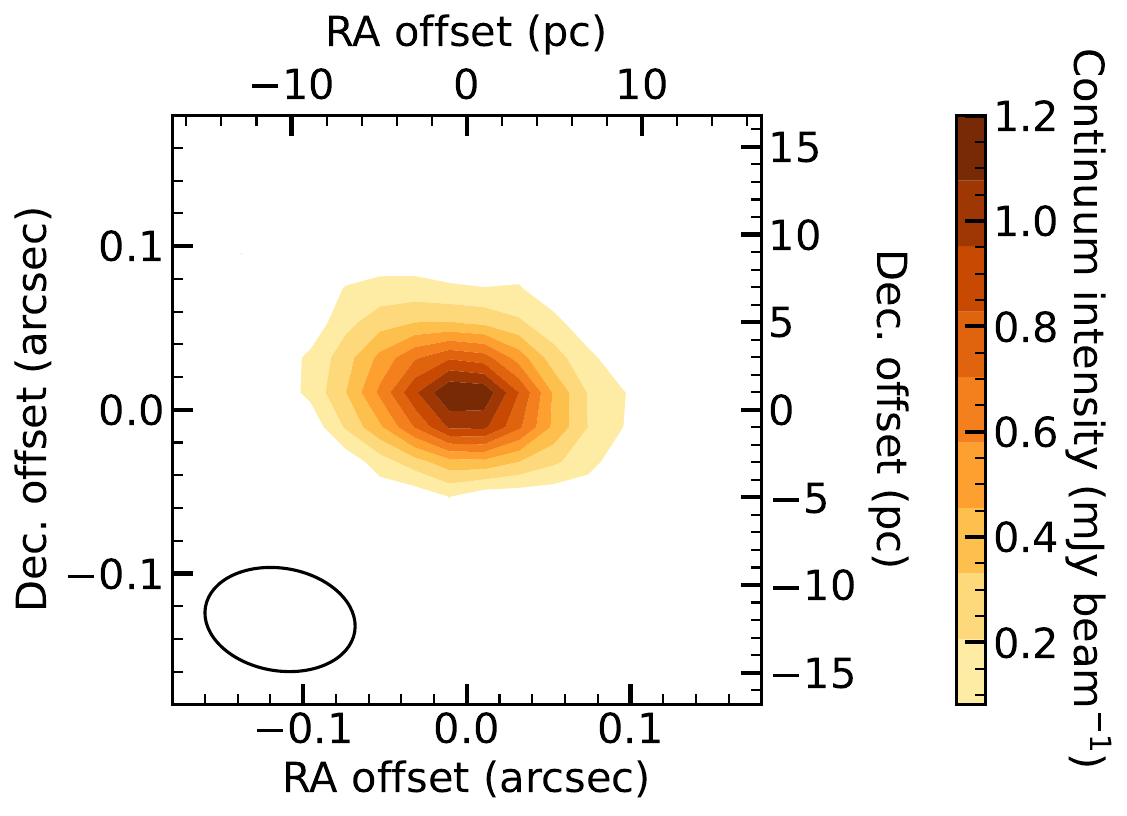}
    \caption{Central region ($0.18$~arcsec~$\times$~$0.18$~arcsec) of the NGC~1387 $1.25$~mm continuum image. Contours are evenly spaced between the peak intensity of $1.20\pm0.03$~mJy~beam$^{-1}$ and 3 times the RMS noise. The synthesised beam is shown in the bottom-left corner as a black open ellipse.}
    \label{fig:continuum}
\end{figure}

An image of the continuum emission of NGC~1387 was created using the $(u,v)$ components from
all four tracks and cleaned with the \textsc{casa} task \texttt{tclean}. The line-free channels of the line spectral windows as well as the continuum spectral windows were used, resulting in a central frequency of $239.4$~GHz ($1.25$~mm). We adopted the multi-frequency synthesis mode, Briggs weighting with a robust parameter of $0.5$ and cleaned the image to a threshold of $3$ times the root-mean-square (RMS) noise of $0.03$~mJy~beam$^{-1}$. We detect a small central source of continuum emission with an integrated flux density of $1.23\pm0.06$~mJy and a peak intensity of $1.20\pm0.03$~mJy~beam$^{-1}$. This continuum flux density is similar to, but inconsistent with, a previous measurement by \citet{WISDOM_XVI} of $1.00\pm0.05$~mJy. A Gaussian fit using the \textsc{casa} task \texttt{imfit} confirms that the deconvolved source size is consistent with a point (i.e.\ an unresolved) source. The continuum image is shown in Fig.~\ref{fig:continuum} and its properties and that of the detected source are summarised in Table~\ref{tab:continuum}.

\begin{table}
    \centering
    \caption{NGC~1387 continuum image and detected continuum point source properties.}
    \begin{tabular}{lc}
        \hline
         Image property & Value \\
         \hline
         Spatial extent (pix) & $1000\times1000$ \\
         Spatial extent (arcsec) & $20\times20$ \\
         Spatial extent (kpc) & $1.9\times1.9$\\
         Pixel scale (arcsec~pix$^{-1}$) & $0.02$ \\
         Pixel scale (pc~pix$^{-1}$) & $1.9$ \\
         RMS noise (mJy beam$^{-1}$) & $0.03$ \\
         Synthesised beam (arcsec) & $0.09\times0.06$ \\
         Synthesised beam (pc) & $8\times6$ \\
         \hline\\
         Source property & Value \\ 
         \hline 
         Right ascension (J2000.0) & $03^{\text{h}}36^{\text{m}}57\dotes0335\pm0\dotes0001$ \\
         Declination (J2000.0) & $-35\degree30\arcmin23\farcs682\pm0\farcs001$ \\
         Integrated flux (mJy) & $1.23\pm0.06$ \\
         \hline  
    \end{tabular}
    \\
    \label{tab:continuum}
\end{table}

\subsection{Line emission}
\label{section:highres data cube}

The $^{12}$CO(2–1) line was similarly imaged using the $(u,v)$ components from all four tracks to create a data cube with high angular resolution. We fitted the same spectral windows used to image the continuum with a fit order of $0$, and this fit was subtracted from the $uv$-plane data using the \textsc{casa} task \texttt{uvcontsub}. These continuum-subtracted $uv$ data were then imaged and cleaned using the cube mode of the \textsc{casa} task \texttt{tclean}. The regions containing emission were cleaned to a threshold equal to $1.5$ times the RMS noise of channels free of any emission in the dirty (i.e.\ not yet cleaned) data cube. For the imaging we used a kinematic local-standard-of-rest velocity reference frame and Briggs weighting with a robust parameter of $0.8$. The conventional robust parameter of $0.5$ yielded very poor signal-to-noise ratios in regions of emission, whereas a parameter of $1.0$ yielded a synthesised beam size that did not sufficiently sample the predicted SoI. A robust parameter of $0.8$ represents a trade-off between the two. We adopted a channel width of $10$~km~s$^{-1}$, again selected to achieve angular (and thus spatial) and velocity resolutions effectively probing the predicted SMBH SoI while providing adequate sensitivity. The resulting data cube has a synthesised beam of $0.13$~arcsec~$\times$~$0.10$~arcsec ($\approx12\times10$~pc$^2$) with a RMS noise of $0.65$~mJy~beam$^{-1}$~channel$^{-1}$. Its properties are summarised in Table~\ref{tab:CO}.

\begin{table}
    \centering
    \caption{CO data cube properties.}
    \begin{tabular}{clc}
        \hline
         Data cube & Image property & Value \\
         \hline
         & Spatial extent (pix) & $1000\times1000$ \\
         & Spatial extent (arcsec) & $30.0\times30.0$ \\
         & Spatial extent (kpc) & $2.8\times2.8$\\
         & Pixel scale (arcsec~pix$^{-1}$) & $0.03$\\
         High-resolution & Pixel scale (pc~pix$^{-1}$) & $2.8$\\
         & Velocity range (km~s$^{-1}$) & $1100$ -- $1400$\\
         & Channel width (km~s$^{-1}$) & $10$\\
         & RMS noise (mJy~beam$^{-1}$~channel$^{-1}$) & $0.65$\\
         & Number of constraints & $347,024$\\
         & Synthesised beam (arcsec) & $0.13\times0.10$\\
         & Synthesised beam (pc) & $12\times10$\\
         \hline 
         
         & Spatial extent (pix) & $300\times300$ \\
         & Spatial extent (arcsec) & $30.0\times30.0$ \\
         & Spatial extent (kpc) & $2.8\times2.8$\\
         & Pixel scale (arcsec~pix$^{-1}$) & $0.1$\\
         Low-resolution & Pixel scale (pc~pix$^{-1}$) & $9.4$\\
         & Velocity range (km~s$^{-1}$) & $1100$ -- $1400$\\
         & Channel width (km~s$^{-1}$) & $10$\\
         & RMS noise (mJy~beam$^{-1}$~channel$^{-1}$) & $0.80$\\
         & Number of constraints & $188,777$ \\
         & Synthesised beam (arcsec) & $0.42\times0.35$\\
         & Synthesised beam (pc) & $39\times33$\\
         \hline
    \end{tabular}
    \label{tab:CO}
\end{table}

\begin{figure*}
  \centering
  \includegraphics[scale=0.45]{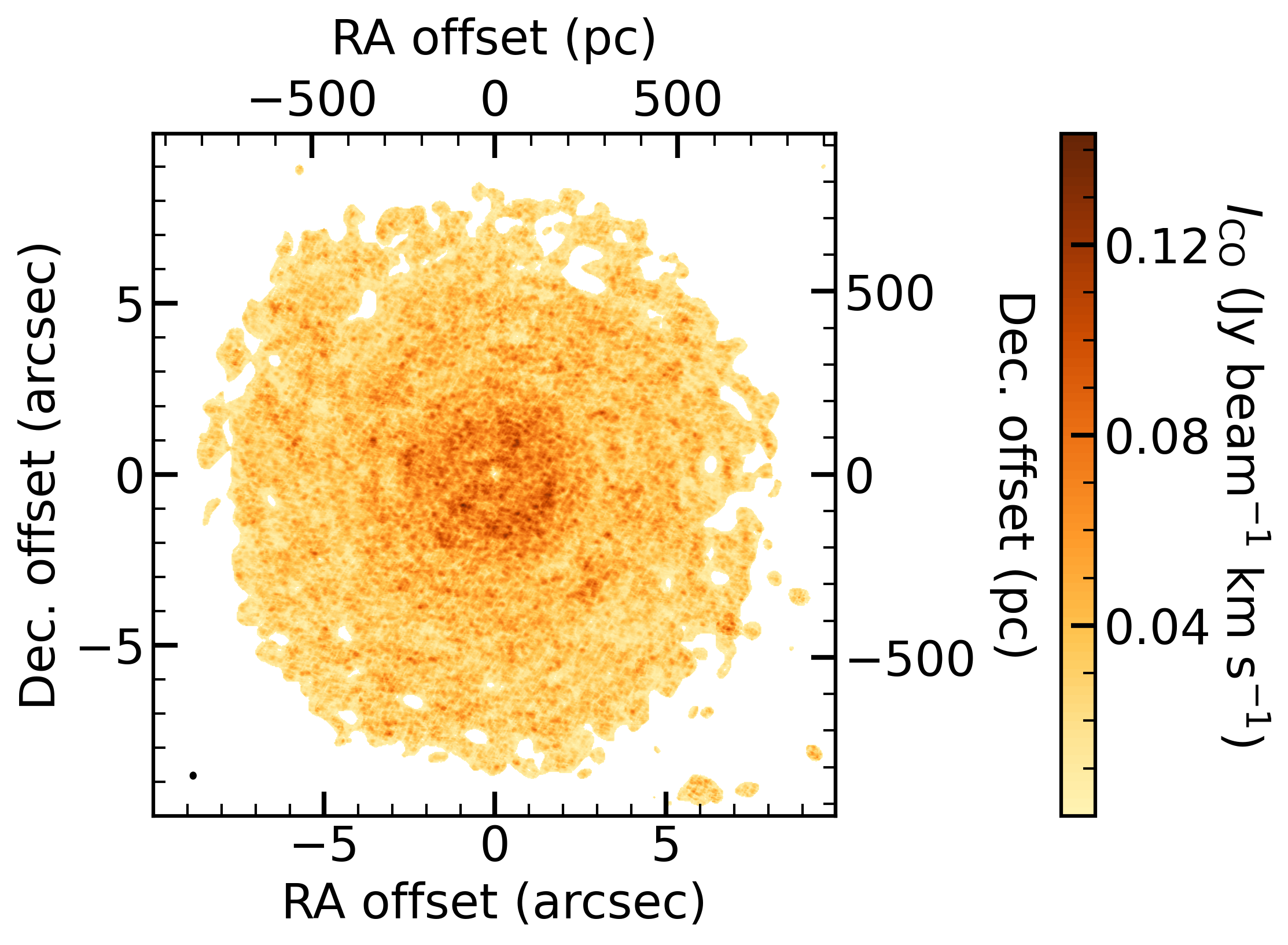}
  \hspace{0.1cm}
  \includegraphics[scale=0.45]{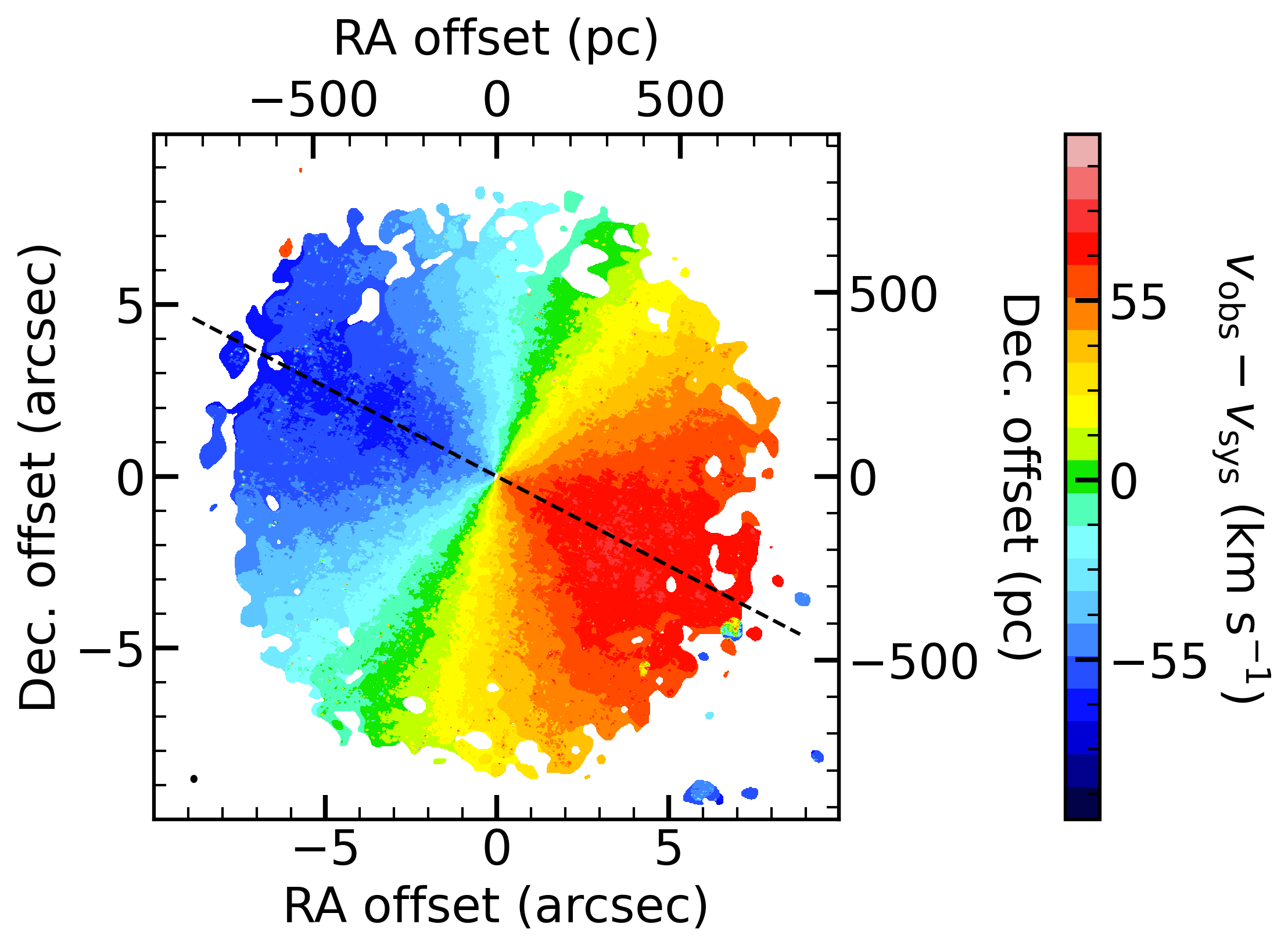}
  \newline
  \hspace*{-1.8cm}
  \includegraphics[scale=0.45]{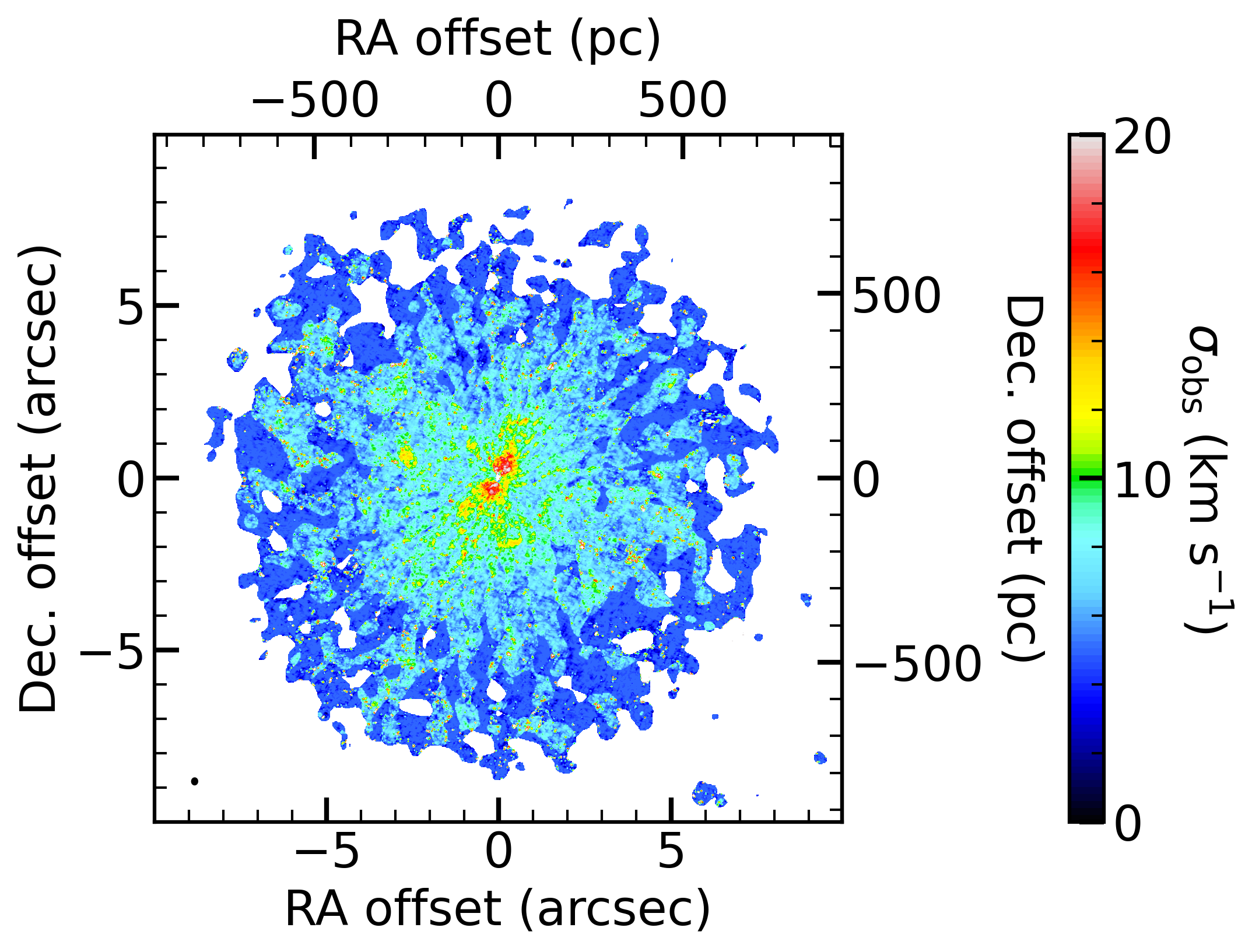}
  %\hspace{0.15cm}
  \includegraphics[scale=0.38]{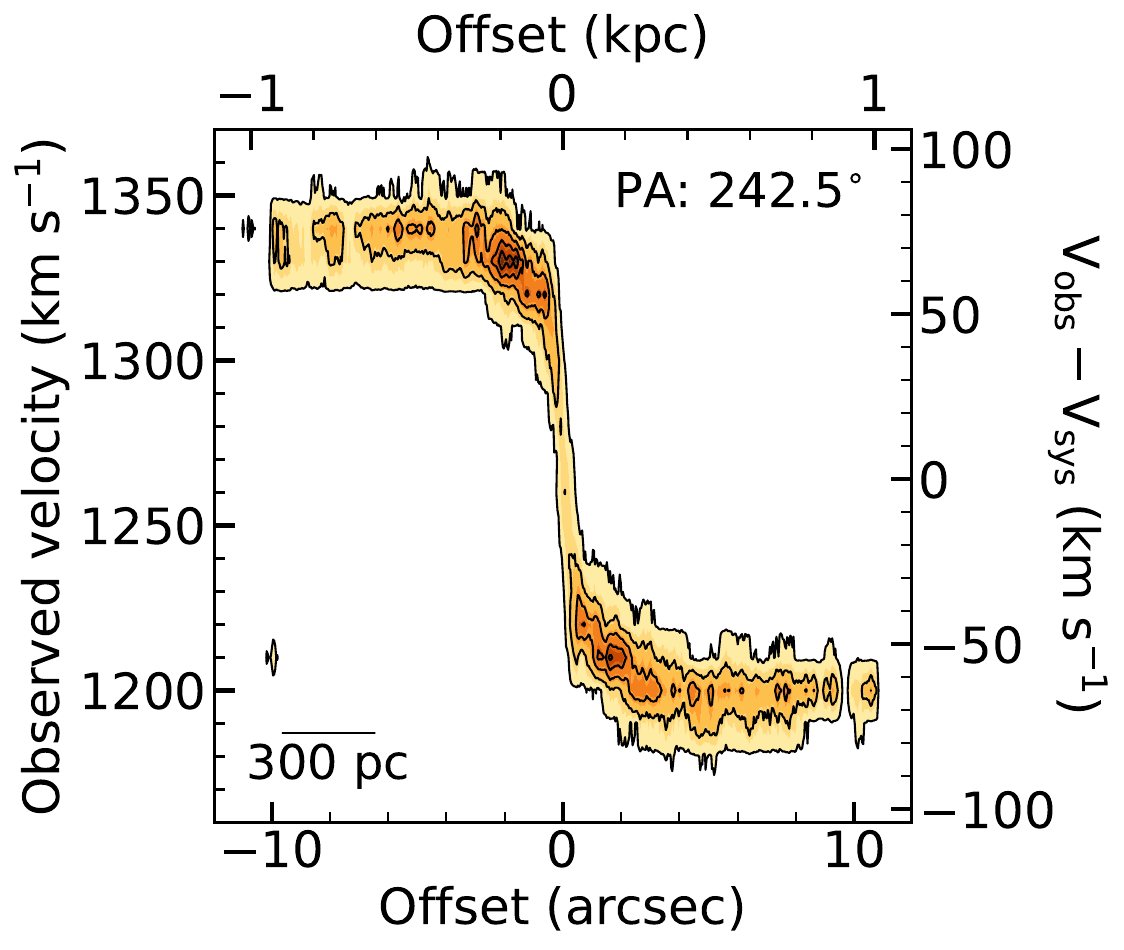}
    \caption{$^{12}$CO(2–1) data products of NGC~1387 derived from our ALMA high-resolution data cube. \textbf{Top-left:} Zeroth-moment (integrated-intensity) map. \textbf{Top-right:} First-moment (intensity-weighted mean line-of-sight velocity) map. The dashed black line shows the kinematic major axis. \textbf{Bottom-left:} Second-moment (intensity-weighted line-of-sight velocity dispersion) map. The synthesised beam is shown in the bottom-left corner of each map as a black filled ellipse. \textbf{Bottom-right:} Kinematic major-axis position-velocity diagram. An error bar is shown in the bottom-left corner, showing the size of the synthesised beam along the kinematic major axis and the channel width.}
  \label{fig:mom012}
\end{figure*}

We used the masked-moment technique of \citet{Dame_2011}, implemented in a modified version of the \textsc{pymakeplots} package\footnote{ \url{https://github.com/TimothyADavis/pymakeplots}}, to create the zeroth (integrated-intensity), first (intensity-weighted mean line-of-sight velocity) and second (intensity-weighted line-of-sight velocity dispersion) moment maps. The primary beam-uncorrected data cube was smoothed with a three-dimensional (3D) Gaussian kernel whose full width at half maximum (FWHM) in each spatial direction is equal to $3$ times the FWHM of the major axis of the synthesised beam, and whose FWHM in the velocity direction is equal to the channel width. All pixels above a clipping threshold equal to $2.5$ times the RMS noise (measured in the unsmoothed primary beam-uncorrected data cube) were selected, creating a mask. This mask was then applied to the unsmoothed primary beam-corrected data cube to create the moment maps. The kinematic major-axis position-velocity diagram (PVD) was constructed by taking a $10$-pixel wide cut through the masked data cube along a position angle (PA) of $242\fdg5$, determined using \textsc{PaFit}\footnote{\url{https://pypi.org/project/pafit/}}. The resulting moment maps and PVD are shown in Fig.~\ref{fig:mom012}.

In NGC~1387 we detect a disc of flocculent, regularly-rotating molecular gas approximately $18$~arcsec ($\approx1.7$~kpc) in diameter, coincident with the central dust disc of the galaxy (see Fig.~\ref{fig:HST}). We also detect a small central depression $\approx0.2$~arcsec ($\approx19$~pc) in radius. This depression is similar in extent to the predicted $R_\text{SoI}$, and thus prevents the innermost parts of the gravitational potential from being probed robustly. This is likely the reason why no clear Keplerian rise of the velocities is present within $R_\text{SoI}$. A SMBH mass measurement is nevertheless still possible in NGC~1387, as the dynamical impact of the SMBH can be traced up to $\approx2$ times the formal SoI given a robust stellar mass model \citep{Davis_2014}. The total flux of NGC~1387 was calculated to be $255\pm3$~Jy~km~s$^{-1}$ in \citet{Liang_2024}, the companion paper to this work. This is the appropriate total flux to use in any follow-up studies, as it has been corrected using the procedure described in \citet{Jorsater_1995} and represents the integrated flux at an infinite cleaning depth.

%%%%%%%%%%%%%%%%%%%%%%%%%%%%%%%%%%%%%%%%%%%%%%%%%%%%%%%%%%%%%%%%%%%%%%%%%%%%%%%%%%%%%%%%%%%%%%%%%%%%
%%%%%%%%%%%%%%%%%%%%%%%%%%%%%%%%%%%%%%%%%%%%%%%%%%%%%%%%%%%%%%%%%%%%%%%%%%%%%%%%%%%%%%%%%%%%%%%%%%%%
\section{Dynamical Modelling}
\label{section:Dynamical Modelling}

\subsection{Modelling method}
\label{subsection:Modelling method}

Dynamical modelling of our data was carried out by fitting a model data cube to the observed molecular gas distribution and kinematics using \textsc{Kinematic Molecular Simulation} (\textsc{KinMS}; \citealt{Davis2013a})\footnote{\url{https://kinms.space}}. This method of SMBH mass determination has been used in all other SMBH mass determinations in the WISDOM series \citep[e.g.][]{Davis2017, Ruffa_2023, Dominiak_2025}. The \textsc{KinMS} method requires a circular velocity curve derived from the stellar light distribution and models of the mass-to-light ratio and the molecular gas distribution. These are then used in creating a model of the molecular gas disc as a collection of point particles, whose line-of-sight projections are calculated to create a model data cube. The instrumental effects are reproduced by spatially convolving this model data cube with the Gaussian synthesised beam, and spatially and spectrally binning it into pixels identical to those of the real data cube.

The \textsc{KinMS} model and ALMA data cubes are then compared in a Bayesian way using the general Gibbs sampler with adaptive stepping \textsc{GAStimator}\footnote{\url{https://github.com/TimothyADavis/GAStimator}}. At first, the Monte Carlo Markov chain (MCMC) algorithm explores the parameter space of all the free parameters of our model. To efficiently explore regions of low probability density and adequately sample regions of high probability density, the step size between each fit can be adjusted until the chain converges. Approximately $10$~per cent of the total number of steps are dedicated to this initial burn-in phase aimed at identifying the convergent chain. The step size distribution is fixed once the MCMC has converged and the algorithm continues to sample the parameter space, producing the final posterior probability distribution.

The parameter space is constrained by a set of prior probability distributions which are typically uniform and thus maximally ignorant. Certain priors are manually defined to ensure a finite converging time whilst others are allowed to cover their entire possible range (e.g.\ a $360\degree$ position angle range). Given these maximally ignorant priors and Gaussian uncertainties on the data cube pixels, the posterior probability distribution of any model is proportional to the log-likelihood function $\ln P\propto-0.5\chi^2$, where $\chi^2$ is defined in the usual manner. The models are fit to a very large number of data points in the 3D data cube, resulting in unrealistically small formal (i.e.\ statistical) uncertainties. However, the dominant sources of uncertainties are in fact systematic, such as our assumption of a perfectly flat axisymmetric disc or our density deprojection. Systematic uncertainties cannot be statistically quantified and require numerical experiments to gauge their impacts (see Section~\ref{subsection:systematic uncertainties}). However, \citet{vandenBosch2009} proposed a heuristic method to address systematic uncertainties by assuming their contribution is similar to that of rescaled statistical uncertainties. \citet{Mitzkus2017} first applied this idea in a Bayesian context, where it involves rescaling the uncertainties of the data cube by a factor of $(2N)^{0.25}$, where $N$ is the number of constraints (i.e.\ the number of pixels with detected emission, as defined by the mask in Section~\ref{section:ALMA Observations} and listed in Table~\ref{tab:CO}). Although not rigorously justified, this method has been adopted in several WISDOM papers and has been shown to yield sensible uncertainties consistent with those derived using a bootstrap approach to systematic effects in \citet[]{WISDOM_IV}.

We implement \textsc{GAStimator} and \textsc{KinMS} using the Python wrapper \textsc{KinMS\_fitter}\footnote{\url{https://github.com/TimothyADavis/KinMS_fitter}}, which provides a simple interface between the two and an easily-accessible integrated library of surface brightness radial profiles.

%%%%%%%%%%%%%%%%%%%%%%%%%%%%%%%%%%%%%%%%%%%%%%%%%%%%%%%%%%%%%%%%%%%%%%%%%%%%%%%%%%%%%%%%%%%%%%%%%%%%
\subsection{Stellar potential}
\label{subsection:stellar potential}

To accurately estimate the SMBH mass from the molecular gas kinematics, we need to quantify the stellar mass contribution to the kinematics. We thus parameterise the intrinsic (i.e.\ `deconvolved') stellar light distribution using a multi-Gaussian expansion (MGE) model \citep{Emsellem1994,Cappellari2002} of a \textit{HST} Wide Field Camera~3 (WFC3) F160W-filter ($H$-band) image. We selected the longest wavelength \textit{HST} image available to minimise dust extinction. 

In more detail, the MGE procedure works by parameterising the two-dimensional (2D) projection of the intrinsic 3D stellar light distribution as a sum of 2D Gaussians, each with a central surface brightness $I^\prime$, a width $\sigma^\prime$ and an axial ratio $q^\prime$. These Gaussian components are first convolved with the spatial point spread function (PSF) of the F160W-filter (generated using \textsc{TinyTim}; \citealt{TinyTim}), itself parameterised using a circularised MGE (see Table~6 of \citealt{Dominiak_2024_MASSIVE}), to reproduce instrumental effects, before being compared and fit to the \textit{HST} image using the \textsc{MgeFit}\footnote{\url{https://pypi.org/project/mgefit/}} \citep{Cappellari2002} algorithm. The surface brightnesses of the Gaussian components can then be converted to luminosity surface densities using the AB magnitude system with a Solar absolute magnitude of $4.60$~mag \citep{Willmer2018} and a zero-point of $25.94$~mag \citep{WFC3}. We adopt a Galactic extinction of $0.006$~mag \citep{Schlafly_2011} from the NASA/IPAC Extragalactic Database to correct for interstellar reddening. 

\begin{table}
	\centering
	\caption{Parameters of the (deconvolved) best-fitting MGE components.}
	\label{tab:MGE components}
	\begin{tabular}{ccc}
	\hline
	$\log\left(\frac{I^\prime} {\text{L}_{\odot,\text{F160W}}~\text{pc}^{−2}}\right)$ & $\log\left(\frac{\sigma^\prime}{\text{arcsec}}\right)$ & $q^\prime$ \\
    (1) & (2) & (3) \\
	\hline
	  $5.068$ & $-0.937$ & $0.98$ \\
	  $4.900$ & $-0.579$ & $0.98$ \\
        $4.617$ & $-0.208$ & $0.98$ \\
        $4.456$ & $\phantom{-}0.233$ & $0.98$ \\
        $4.006$ & $\phantom{-}0.607$ & $0.98$ \\
        $3.463$ & $\phantom{-}0.849$ & $0.98$ \\
        $2.748$ & $\phantom{-}1.214$ & $0.98$ \\
        $2.414$ & $\phantom{-}1.725$ & $0.98$ \\
	\hline
    \end{tabular}
    \begin{tablenotes}
          \item \textit{Notes.} MGE Gaussian components. (1) Central surface brightness. (2) Standard deviation. (3) Axial ratio.
    \end{tablenotes}
\end{table}

\begin{figure*}
    \centering
    \includegraphics[scale = 0.55, trim = {0.5cm 0 0 0}]{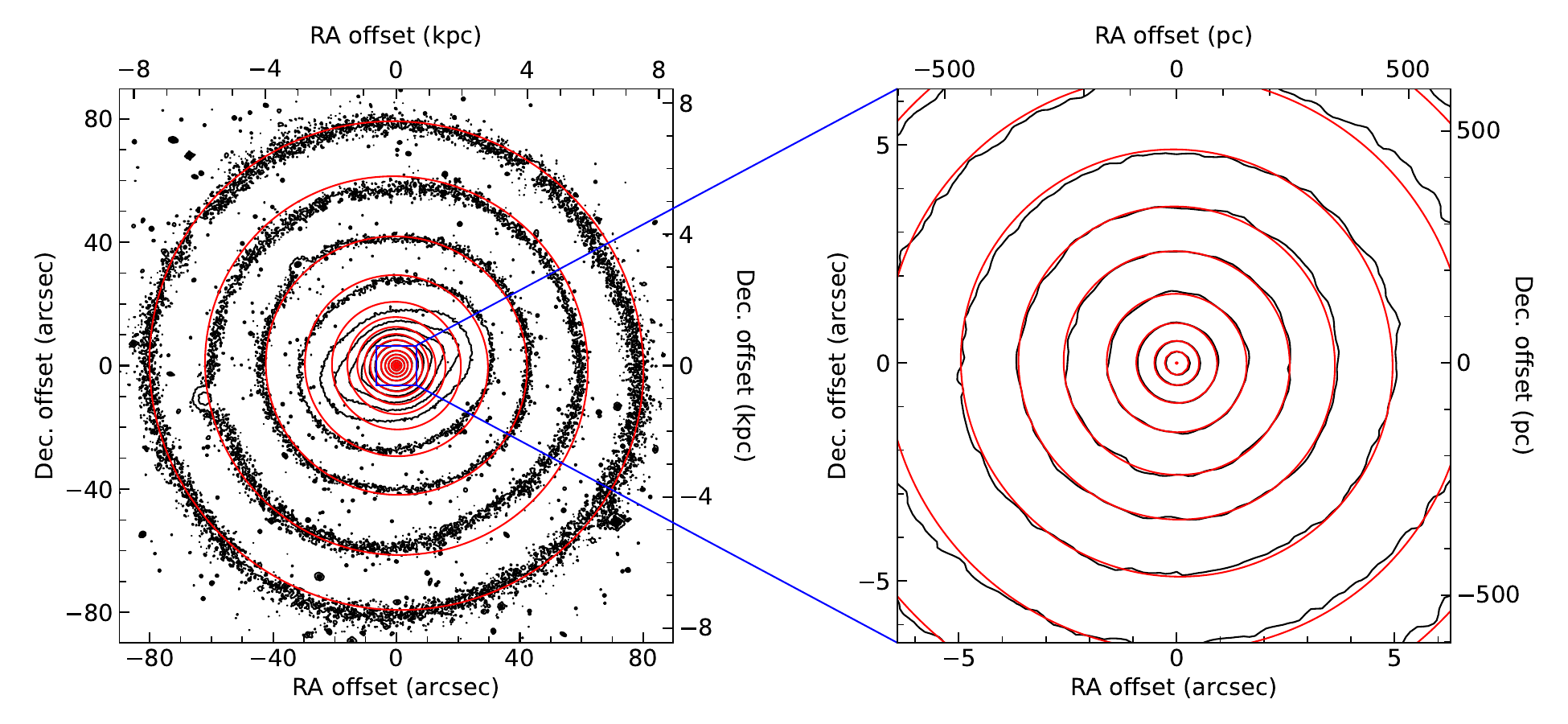}
    \caption{\textit{HST} WFC3 F160W-filter ($H$-band) image of NGC~1387 (black contours), overlaid with our best-fitting MGE model (red contours). \textbf{Left:} Full extent of the galaxy, showing the galaxy bulge, bar and extended stellar disc. \textbf{Right:} Central region only.}
    \label{fig:MGE contours}
\end{figure*}

Our MGE modelling relies on the assumption of axisymmetry, however NGC~1387 has a non-axisymmetric bar feature. MGE modelling can still be employed despite this, as the dominant disc is axisymmetric and the bar can be treated as a first-order perturbation of an otherwise axisymmetric system. In such a case, a standard procedure is to fit a single Gaussian axial ratio in the outer regions of the galaxy (far away from the bar) and to fix the axial ratios of all the Gaussians to this, even across the barred region (see \citealt{SAURON_XIV} for further details). Although this method may produce a poorer fit within the barred region, it provides a much better axisymmetric approximation of the gravitational potential of the disc, for use in our axisymmetric models. If we were to incorrectly interpret the bar as a nearly edge-on axisymmetric disc embedded into a spherical spheroid, and fit its detailed surface brightness, the deprojected intrinsic density would completely misrepresent the true density distribution. In any case, this approximation has little impact on our model, as the gas disc we model is limited to the central regions, where the influence of the bar is small. The roundness of the outer isophotes makes it difficult to constrain the (photometric) PA of the disc line of nodes, so we determine the (kinematic) PA from the molecular gas data using \textsc{PaFit}. We then orient the MGE along this kinematic PA of $242\fdg5$. For NGC~1387, we derive an axial ratio of $0.98$ by inspecting the outer regions of the \textit{HST} image and assuming an infinitely-thin disc. 

The parameters of the (`deconvolved') best-fitting Gaussians (i.e.\ those Gaussians best representing the 2D projection of the intrinsic 3D stellar light distribution, without instrumental effects) are listed in Table~\ref{tab:MGE components} in physical units. The fit is shown in Fig.~\ref{fig:MGE contours} overlaid on the data. There is rather uniform dust extinction across the entire extent of the galaxy, so we have opted not to mask any part of the disc. The fit is superficially poor in the barred region of the galaxy (see the left panel of Fig.~\ref{fig:MGE contours}), but the resulting MGE model is clearly an excellent fit at small radii (see the right panel of Fig.~\ref{fig:MGE contours}), the most important region to constrain the SMBH mass, as well as the outermost radii unaffected by the bar.

Alongside the free model parameters of inclination $i$, constant stellar $M/L$ and SMBH mass, we input our 2D MGE luminosity surface density model into the \texttt{mge\_vcirc} routine of the \textsc{Jeans Anisotropic Modelling} Python package (\textsc{JamPy}\footnote{\url{https://pypi.org/project/jampy/}}; \citealt{Cappellari_2020}) to calculate the circular velocity curve. Here, using the inclination $i$, our 2D MGE luminosity surface density model is analytically deprojected into a 3D luminosity volume density distribution. The deprojection inferring the intrinsic density from the observed surface brightness is mathematically non unique and the degeneracy becomes particularly severe at low inclination \citep{Rybicki1987}, as in our case. The MGE formalism \citep{Cappellari2002} provides one possible deprojection that is consistent with the observed surface brightness and, in this case, has a constant axial ratio. However, the deprojection is only defined if $\cos^2i<{q^\prime}^2$ for all Gaussians, so that the flattest Gaussian of an MGE model dictates the minimum possible inclination for which an MGE model can be used in a dynamical model. As all of our Gaussians have a fixed axial ratio $q^\prime=0.98$, the minimum inclination for which we are able to deproject is $i=11.5\degree$, which is sufficient for our purposes. The resulting 3D luminosity volume density distribution is then converted into a 3D mass volume density distribution by simply multiplying each MGE component by the constant stellar $M/L$, that is also a free parameter of our model. This stellar mass model and our SMBH mass are then used to calculate the associated circular velocity curve, which is input into \textsc{KinMS}.

%%%%%%%%%%%%%%%%%%%%%%%%%%%%%%%%%%%%%%%%%%%%%%%%%%%%%%%%%%%%%%%%%%%%%%%%%%%%%%%%%%%%%%%%%%%%%%%%%%%
\subsection{Molecular gas distribution}
\label{subsection:Molecular gas distribution}

To create the model data cube, \textsc{KinMS} also requires a model of the observed gas distribution. For this, we can use the observed but deconvolved molecular gas emission. We implement this using the custom-made \textsc{KinMS} plugin \textsc{SkySampler}\footnote{\url{https://github.com/Mark-D-Smith/KinMS-skySampler}}, described by \citet{WISDOM_IV}. \textsc{SkySampler} first obtains the kinematics-independent intrinsic distribution of the gas by integrating the point source model (i.e.\ the clean components yielded by the \textsc{casa} task \texttt{tclean}) of the gas distribution along the velocity axis. This distribution is then sampled with $5\times10^6$ particles. Assuming a certain position angle and inclination of the galaxy, the particle positions generated by \textsc{SkySampler} can then be deprojected into their intrinsic positions in the galaxy plane using the \textsc{KinMS\_fitter} task \texttt{transformClouds}.

%%%%%%%%%%%%%%%%%%%%%%%%%%%%%%%%%%%%%%%%%%%%%%%%%%%%%%%%%%%%%%%%%%%%%%%%%%%%%%%%%%%%%%%%%%%%%%%%%%%
\subsection{Low-resolution cube and inclination prior}
\label{subsection: inclination prior}

When using \textsc{SkySampler}, the inclination of the galaxy remained poorly constrained ($\pm5\degree$ uncertainty), most likely due to the the face-on orientation of the galaxy and the sparsity of gas in the outer regions of the disc. As \textsc{SkySampler} models the gas distribution as an array of cloudlets rather than a cohesive and continuous disc, it effectively removes constraints that arise from the disc geometry. Instead, only the observed particle velocities are left, which are largely degenerate with the inclination. As such, \textsc{SkySampler} is known to struggle to constrain the inclination. Nevertheless, due to the flocculence of the molecular gas, using \textsc{SkySampler} is the only way we can reliably reproduce the surface brightness of NGC~1387 (an analytic surface brightness distribution would have too many parameters).

\begin{figure}
    \centering
    \includegraphics[scale = 0.44]{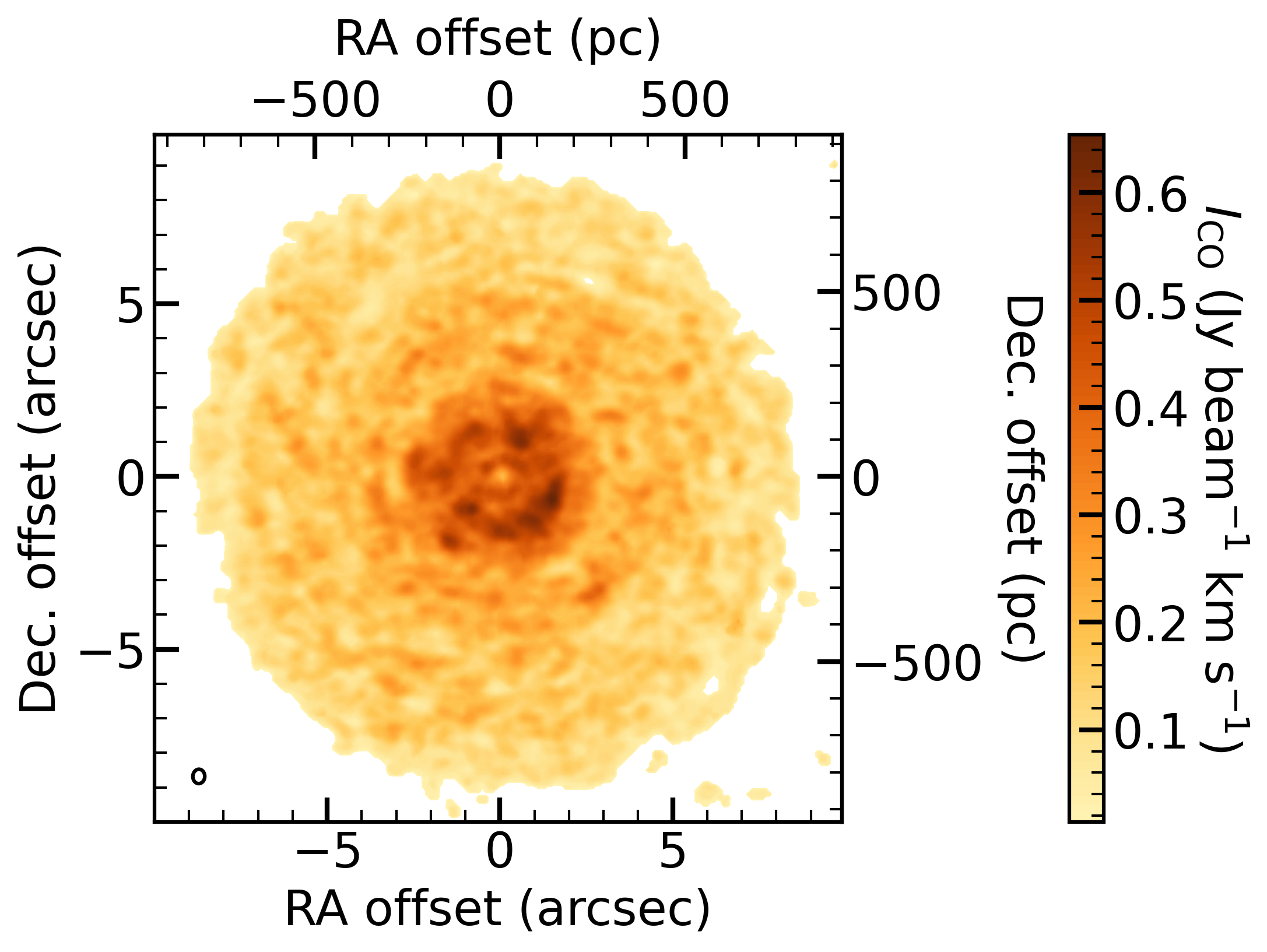}
    \caption{$^{12}$CO(2–1) zeroth-moment (integrated-intensity) map of NGC~1387 derived from our ALMA low-resolution data cube. The synthesised beam is shown in the bottom-left corner as a black open ellipse.}
    \label{fig:low res cube}
\end{figure}

We overcome this problem by creating a second, lower-resolution data cube. The same continuum-subtracted $uv$ data as for the high-resolution cube were imaged and cleaned in an analogous manner, but this time using Briggs weighting with a robust parameter of $2.0$, corresponding to natural weighting and providing a decreased angular resolution but increased sensitivity. Additionally, we used the \texttt{uvtaper} option to apply a FWHM taper of $600$~kilolambda ($\approx780$~m), to further degrade the angular and thus spatial resolution (to $\approx3$ times that of our high-resolution data cube) and improve the sensitivity of the data cube. As with the high-resolution data cube, we adopt a channel width of $10$~km~s$^{-1}$. The resulting data cube has a synthesised beam of $0.42$~arcsec~$\times$~$0.35$~arcsec ($\approx39\times33$~pc$^2$) with a RMS noise of $0.80$~mJy~beam$^{-1}$~channel$^{-1}$, and its properties are also summarised in Table~\ref{tab:CO}. Moment maps were again created, using the same method as described in Section~\ref{section:highres data cube}, but in this case the Gaussian kernel's FWHM in each spatial direction was equal to the FWHM of the major axis of the synthesised beam. The mask was obtained by clipping the smoothed primary beam-uncorrected data cube to a threshold of $5$ times its RMS noise. The resulting zeroth-moment map is shown in Fig.~\ref{fig:low res cube}.

First, by degrading the angular resolution, we ensure higher signal-to-noise ratios in emission regions, particularly in the outer parts of the disc, making the determination of the boundaries of the molecular gas disc much easier. Second, due to the lower resolution, the molecular gas disc appears smoother and more regular, allowing to model it assuming an infinitely-thin axisymmetric disc with an exponential radial surface brightness profile. This eliminates the need to use \textsc{SkySampler} in this step, allowing us to use this lower-resolution cube to more reliably constrain the inclination prior to fitting the high-resolution cube. Due to the small central depression in both the higher- and the lower-resolution data cube, we also allowed for the possibility of an inner truncation in the radial surface brightness profile. The molecular gas distribution is thus parameterised by
\begin{equation}
    I(R)\propto 
    \begin{cases}
      0 & R\leq R_{\text{hole}}\\
      e^{-R/R_0} & R>R_{\text{hole}}\\
    \end{cases}   
    ,
\label{eq:surface density}
\end{equation}
where $R$ is the galactocentric radius, $R_0$ the exponential scale length and $R_{\text{hole}}$ the truncation radius, the latter two being free parameters of our model. \review{As NGC~1387 appears to have a depression rather than an outright hole, if we were modelling the SMBH mass in this manner it might be more appropriate to consider models with a partially-filled inner region. However, this profile is only used for the purpose of modelling the low-resolution cube to obtain an inclination prior. It is therefore not essential to accurately reproduce the innermost region of the galaxy. Instead, it is more important to accurately model the whole extent of the molecular gas disc. The exact choice of innermost profile is therefore unlikely to affect the inclination (prior) determination.}

We thus create a \textsc{KinMS} model in the standard manner, using the stellar potential described in Section~\ref{subsection:stellar potential} and the molecular gas distribution described above. This model has a total of $11$ free parameters: SMBH mass ($M_\text{BH}$), spatially-constant stellar mass-to-light ratio ($M/L$), molecular gas disc scale length ($R_0$), truncation radius ($R_{\text{hole}}$), integrated flux density and spatially-constant velocity dispersion ($\sigma_\text{gas}$), as well as the "nuisance" parameters describing the disc geometry position angle, inclination ($i$), systemic velocity ($V_\text{sys}$) and centre position (offsets in right ascension and declination from the image phase centre). The priors of all the parameters are uniform, apart from that of the SMBH mass which is uniform in the logarithm due to its large dynamic range.

\begin{figure}
    \centering
    \includegraphics[scale = 0.5, trim = {1.0cm 0 0cm 0}]{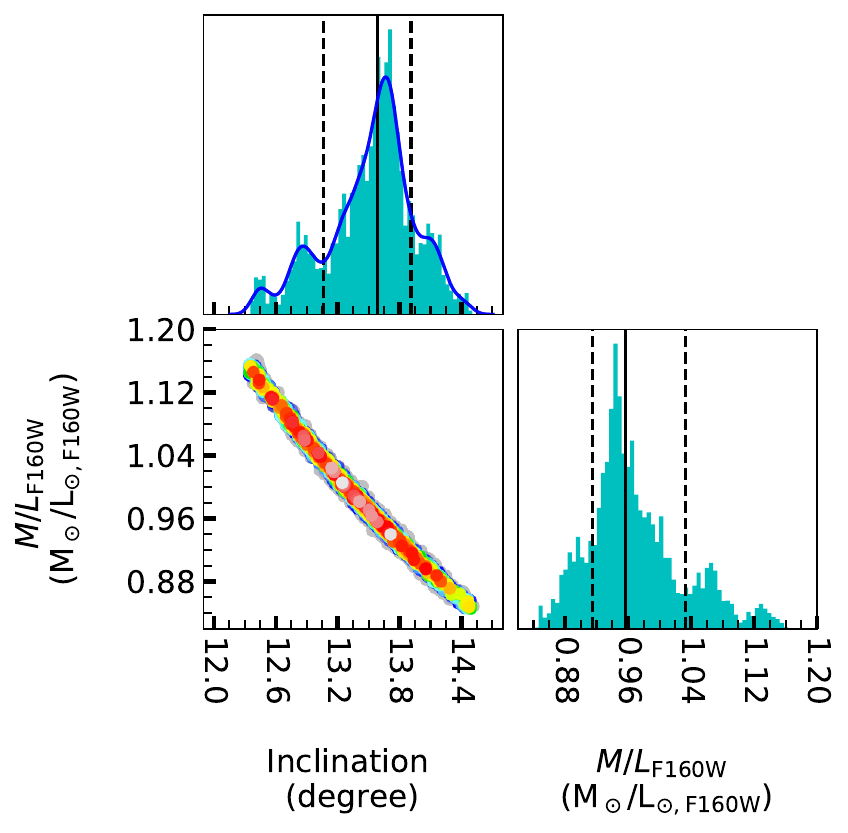}
    \caption{Corner plot of NGC~1387, showing the covariance between $i$ and $M/L$ from the first modelling step. Each data point is a realisation of our model, colour-coded to show the relative log-likelihood of that realisation, with white data points most likely and blue data points least likely. Coloured points show models with $\Delta\chi^2<\sqrt{2N}$ of the best-fitting model; grey data points show all remaining models. The histograms show the 1D marginalised posterior probability distribution of each model parameter, where the black solid line marks the median and the two black dashed lines the $68\%$ confidence interval. The kernel density estimation of the inclination posterior probability distribution is shown as a solid blue line.}
    \label{fig:cornerPlot_inc}
\end{figure}

As a part of the MCMC chain, every free parameter of our model undergoes a marginalisation process. Figure~\ref{fig:cornerPlot_inc} shows the probability distributions of $i$ against $M/L$. Each data point is colour-coded by the log-likelihood of that model, with white data points indicating the highest likelihood and blue data points the lowest. Additionally, one-dimensional (1D) marginalisations, represented as histograms, show the posterior probability distributions of the two parameters. The convergence of the chain is indicated by the fact that both histograms \review{have a well-defined large central peak which tapers out towards the edges.}

The MCMC chain in that first stage modelling had $200,000$ steps, yielding a best-fitting inclination $i=13.6\degree^{+0.9}_{-1.2}$. Figure~\ref{fig:cornerPlot_inc} reveals a very strong degeneracy between $i$ and $M/L$, but this is to be expected given the small inclination. For a dynamical mass measurement based on rotational velocities, the enclosed mass $M_\text{tot}$ is related to the rotation velocity $v_\text{rot}$ and thus the observed velocity $v_\text{obs}$ as $M_\text{tot}(R)\propto v_\text{rot}^2(R)R\propto v_\text{obs}^2(R)R/\sin^2i\propto\sin^{-2}i$, and $\sin i$ varies rapidly at small inclinations (see \citealt{WISDOM_IV} for further discussion of this issue). Nevertheless, the MCMC chain of this initial stage of modelling converges without issue and the posterior probability distribution of the SMBH mass is Gaussian in shape, already strongly suggesting the presence of a central SMBH.

To model the high-resolution data cube with \textsc{SkySampler}, we adopt the posterior probability distribution of the inclination (obtained from our modelling of the low-resolution data cube without \textsc{SkySampler}) as a prior. As the inclination posterior probability distribution is comprised of a collection of discrete points, in practice we adopt a smooth approximation of that distribution as the prior. In other works that implement this method \citep[e.g.][]{WISDOM_XXII}, this was achieved by fitting a parametric function to the posterior probability distribution, normally a Gaussian. However, in the case of NGC~1387, the inclination posterior probability distribution cannot be easily approximated using a parametric function, as it has an irregular shape with three smaller peaks around a central dominant one (see Fig.~\ref{fig:cornerPlot_inc}). We therefore instead choose to use a kernel density estimation (KDE), whereby the histogram is smoothed by placing a Gaussian kernel at each data point in the posterior distribution. The kernels are then summed to create a continuous function, which is normalised so that the area under the curve is equal to $1$, to be used as the prior. To achieve this, we used the \texttt{neighbors.KernelDensity} task, part of the \textsc{scikit-learn} library, with a kernel width of $0.07$, which provides sufficient smoothing of the histogram data but does not erase the overall shape of the posterior probability distribution. The kernel density estimation is overlaid on the inclination posterior probability distribution in Fig.~\ref{fig:cornerPlot_inc}.

%%%%%%%%%%%%%%%%%%%%%%%%%%%%%%%%%%%%%%%%%%%%%%%%%%%%%%%%%%%%%%%%%%%%%%%%%%%%%%%%%%%%%%%%%%%%%%%%%%%
\subsection{Results}
\label{subsection:results}

Our final results are generated by comparing the high-resolution data cube described in Section~\ref{section:highres data cube} to a \textsc{KinMS} model created using the stellar potential described in Section~\ref{subsection:stellar potential} and \textsc{SkySampler}, together with the inclination prior derived in Section~\ref{subsection: inclination prior}. This model has a total of $7$ free parameters: SMBH mass, spatially-constant stellar mass-to-light ratio, molecular gas disc integrated flux density and spatially-constant molecular gas velocity dispersion, and the nuisance parameters position angle, inclination and systemic velocity. We have opted to fix the centre position that of the detected continuum emission, as the latter is consistent with the kinematic centre determined in the previous modelling step (low-resolution data cube) and the reduction of the model dimensionality leads to a decrease of the computing time. Additionally, given the presence of the central depression in the molecular gas and the associated lack of constraints from the disc geometry, \textsc{SkySampler} struggled to accurately determine the kinematic centre. The SMBH mass prior is again uniform in the logarithm, the inclination prior is set by the previous low-resolution data cube modelling (see Section~\ref{subsection: inclination prior}), and all remaining parameters again have uniform priors. The model parameters and their priors (i.e.\ search ranges), best-fitting values and $1\sigma$ ($68.3$~per cent) and $3\sigma$ ($99.7$~per cent) uncertainties are listed in Table~\ref{tab:Best Fit}.

\begin{table*}
    \begin{threeparttable}
	\centering
	\caption{Best-fitting model parameters and associated uncertainties}
	\label{tab:Best Fit}
	\begin{tabular}{lrcrcrr}
		\hline
	Parameter & & Prior & & Best fit & $1\sigma$           uncertainty%\tnote{1} 
        & $3\sigma$ uncertainty \\
        \hline
        \textbf{Mass model} \\
        $\log(M_\text{BH}/\text{M}_\odot)$ & $7$ & $\rightarrow$ & $11$ & $8.04$ & $-0.23,+0.18$ & $-0.87,+0.41$ \\
        Stellar $M/L_{\text{F160W}}$ (M$_\odot$/L$_{\odot,\text{F160W}}$) & $0.0$ & $\rightarrow$ & $5.0$ & $0.90$ & $-0.14, +0.16$ & $-0.35,+0.44$ \\
        \\
        \textbf{Molecular gas disc} \\		
        Velocity dispersion (km~s$^{-1}$) & $0.0$ & $\rightarrow$ & $50.0$ & $6.0$ & $\pm0.3$ & $-0.8,+0.9$ \\
        Integrated intensity (Jy~km~s$^{-1}$) & $0$ & $\rightarrow$ & $300$ & $264$ & $\pm9$ & $-26, +25$ \\
        
        \\ 
        \textbf{Nuisance parameters} \\
        Inclination (degree) & $11.5$ & $\rightarrow$ & $20.0$ & $14.2$ & $-1.1, +1.3$ & $-2.6, +4.1$ \\
        Position angle (degree) & $0.0 $ & $\rightarrow$ & $359.9$ & $243.9$ & $\pm0.3$ & $-1.0, +0.9$ \\
        Systemic velocity (km~s$^{-1}$) & $1100$ & $\rightarrow$ & $1400$ & $1267.2$ & $\pm0.3$ & $-0.7, +0.8$ \\
        \hline
	\end{tabular}
 %\\
    \begin{tablenotes}[flushleft]
          \item \textit{Note.} The centre position is fixed to that of the unresolved continuum source, $03^{\text{h}}36^{\text{m}}57\dotes 0335$, $-35\degree 30\arcmin 23\dotarc 682$ (J2000.0).
    \end{tablenotes}
    \end{threeparttable}
\end{table*}

\begin{figure*}
    \centering
    \includegraphics[scale=0.5, trim={1.0cm 0 0cm 0}]{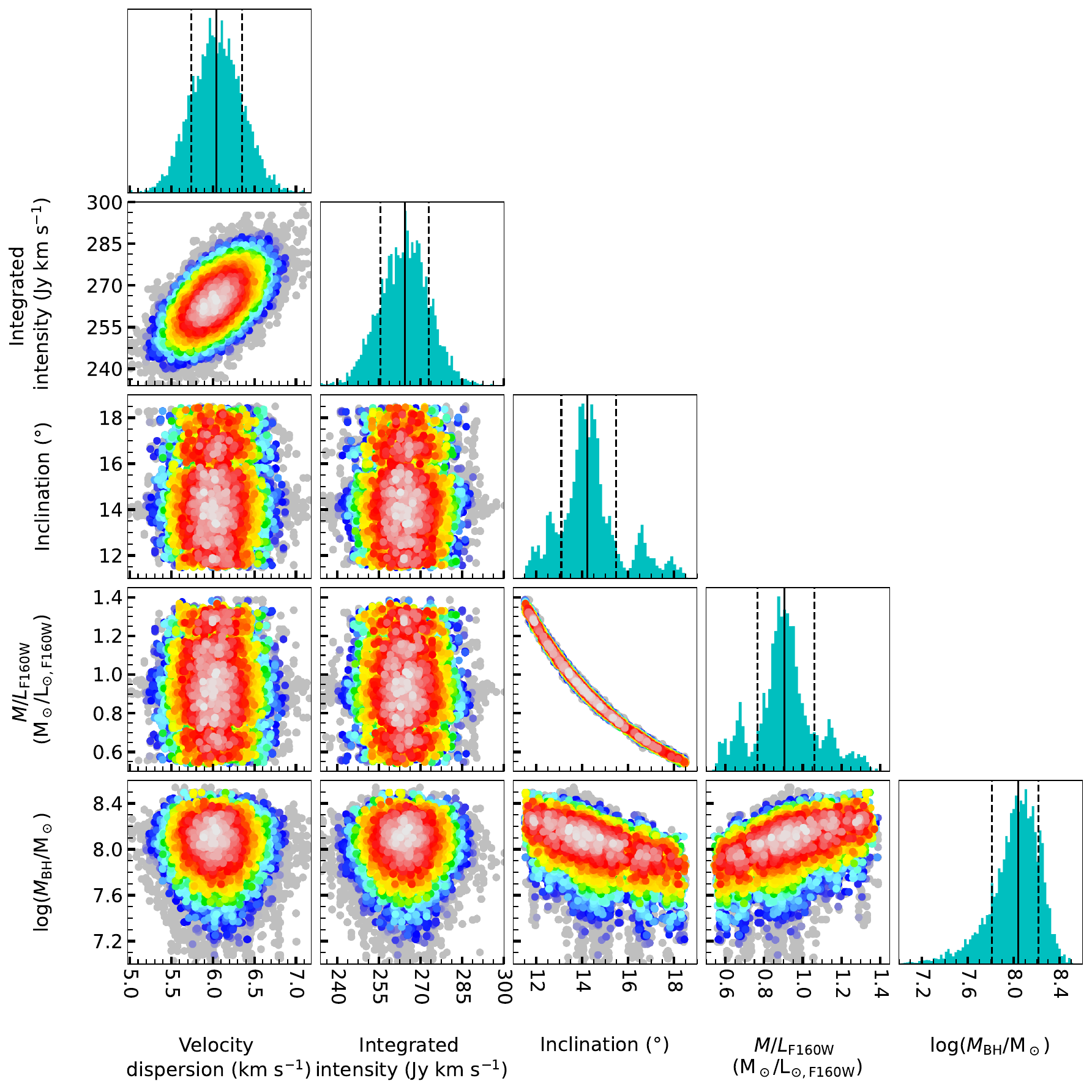}
    \caption{As Fig.~\ref{fig:cornerPlot_inc}, but for selected (primarily non-nuisance) model parameters from the second and final modelling step.}
    \label{fig:cornerPlot}
\end{figure*}

The final MCMC chain in this second stage of our modelling had $150,000$ steps. Figure~\ref{fig:cornerPlot} shows the posterior probability distribution of each non-nuisance model parameter and the inclination against all other parameters. It unequivocally shows that there is a massive dark object at the centre of NGC~1387, with a best-fitting mass of $1.10^{+1.71}_{-0.95}\times10^8$~M$_\odot$, where here and throughout the uncertainties are stated at the $3\sigma$ ($99.7$~per cent) confidence level. The best-fitting $M/L$ in the F160W-filter is $0.90^{+0.44}_{-0.35}$~M$_\odot$/L$_{\odot,\text{F160W}}$. As expected due to the face on nature of the galaxy, Fig.~\ref{fig:cornerPlot} shows a very strong degeneracy between $M/L$ and inclination and a weaker degeneracy between $M_{\text{BH}}$ and inclination. Figure~\ref{fig:cornerPlot} also reveals a positive correlation between $M_{\text{BH}}$ and $M/L$ \citep[see also][]{WISDOM_IV}, as the effects of inclination are sufficiently strong in this face-on system to overwhelm the usual degeneracy arising from the conservation of total dynamical mass.

\begin{figure*}
    \centering\includegraphics[scale=0.58, trim={4cm 0 3cm 0}]{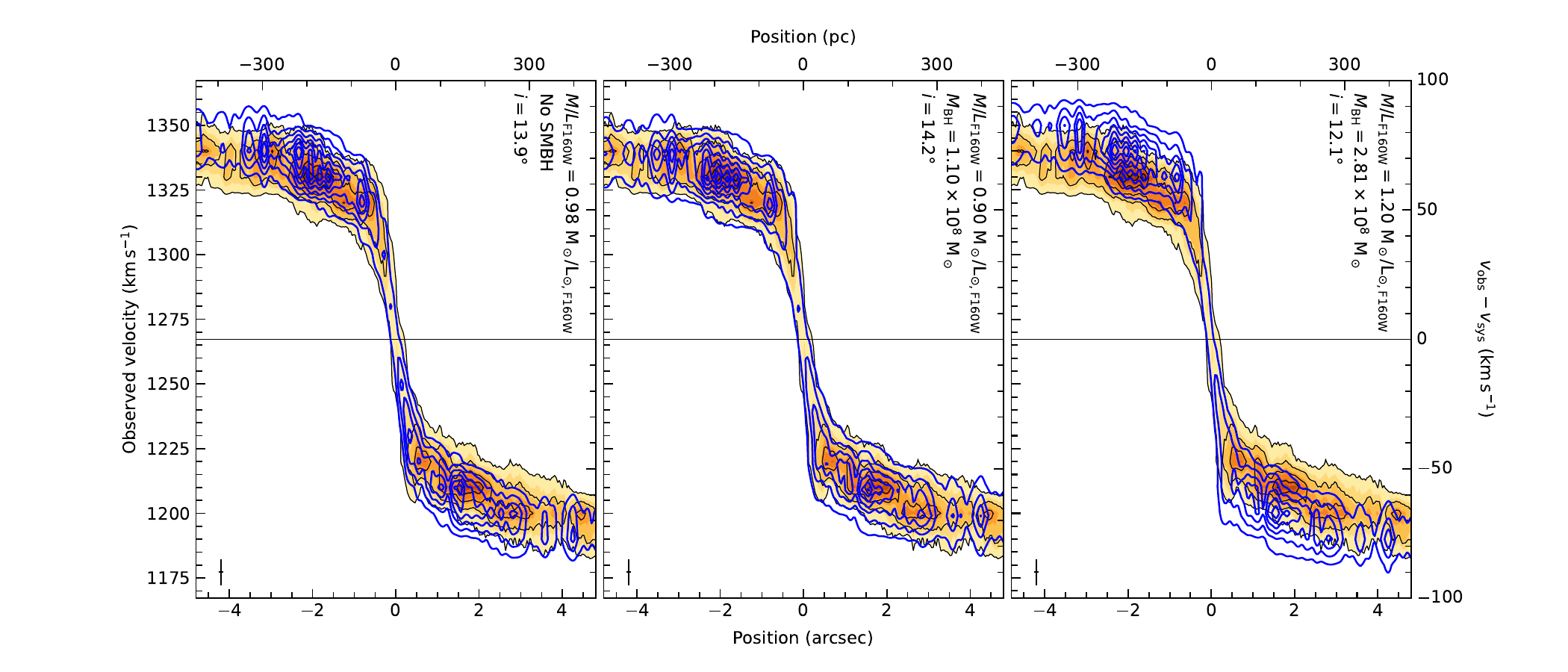}
    \caption{Inner part (radii $R\lesssim5.0$~arcsec) of the observed kinematic major-axis PVD of NGC~1387 (orange scale with black contours), overlaid with the best-fitting no SMBH (left), free SMBH mass (centre) and overly-massive SMBH (right) model (blue contours), respectively. The SMBH mass, $M/L$ and inclination of each model are listed in the top-right corner of each panel. An error bar is shown in the bottom-left corner of each panel, showing the size of the synthesised beam along the kinematic major axis and the channel width. The need for a central dark mass to fully account for the gas kinematics at all radii is clear.}
    \label{fig:PVD_NGC1387}
\end{figure*}

The quality of our best-fitting model can be judged by overlaying it over the kinematic major-axis PVD, as done in the central panel of Fig.~\ref{fig:PVD_NGC1387}. For comparison, we also generate and overlay two other models, one with no SMBH (left panel of Fig.~\ref{fig:PVD_NGC1387}) and one with an overly-massive SMBH (by $\approx3\sigma$ more massive than the best-fitting SMBH; right panel of Fig.~\ref{fig:PVD_NGC1387}), allowing all parameters other than the SMBH mass to vary. As expected, the model with no SMBH tries to compensate for the lack of a central SMBH by adopting a higher $M/L$ to account for the high rotation velocities at small radii. However, the model with no SMBH is not able to fully account for the central velocities in an adequate manner and provides a poor fit in those central regions. Also due to the higher $M/L$, the the fit is noticeably poorer at larger radii. It would also be typically expected that the best-fitting model with an overly-massive SMBH would be forced towards a smaller $M/L$, however in this case the effect is opposite. As shown by Fig.~\ref{fig:cornerPlot}, a higher $M_\text{BH}$ tends towards a lower inclination which in turn is correlated with a higher $M/L$. This is exactly what happens in the model with an overly-massive SMBH which drives the best-fitting inclination down to $i=12.1\degree$ which in turn results in a higher $M/L$ than that of our best-fitting model. However, due to the overly massive SMBH, the model over-shoots the central velocities despite the lower inclination, resulting again in a very poor fit.

\review{We can also quantify the goodness-of-fit of each model in a more rigorous statistical manner. We do so by looking at the reduced $\chi^2$ ($\chi^2_\text{red}$) statistic of each model. Indeed, the $\chi^2$ statistic is most often used to compare models against data, but we can also use it in a heuristic way to measure the relative fit quality of model against model. The model with the best-fitting SMBH has $\chi^2_\text{red}=1.117$, whereas the no SMBH model has $\chi^2_\text{red}=1.134$ (see Table~\ref{table:chi2 values}). Whilst the difference between these two models is ostensibly small with $\Delta\chi_\text{red}^2=0.017$, the expected standard deviation of the $\chi^2_\text{red}$ distribution is $\sqrt{2/N}=0.0024$ for the $N=347,024$ pixels of our high-resolution cube. This indicates that the $\chi^2_\text{red}$ difference measured significantly exceeds the random fluctuations expected from the $\chi^2_\text{red}$ statistic, and in turn that we can reliably differentiate between those two models \citep{Andrae_2010}. It is thus clear that the model with a non-zero SMBH mass is quantitatively preferred.}

\begin{table}
	\centering
	\caption{\review{Goodness-of-fit statistics of the models discussed.}}
	\label{table:chi2 values}
	\begin{tabular}{lccc}
	\hline
	Model & $\chi^2_\text{red}$ \\%& k &  Rescaled $\chi^2$ \\
    (1) & (2) \\% & (3) & (4) \\
        \hline
        Model with no SMBH & $1.134$ \\% & $6$ & $472.3$\\
        Model with best-fitting SMBH & $1.117$ \\% & $7$ & $465.1$ \\
        Model with over-massive SMBH & $1.125$ \\% & $7$ & $468.7$ \\
        Model with dust correction & $1.116$ \\% & $7$ & $464.9$ \\
        Model with linearly-varying $M/L$ & $1.108$ \\% & $8$ & $461.6$ \\
	\hline
    \end{tabular}
\end{table}

%%%%%%%%%%%%%%%%%%%%%%%%%%%%%%%%%%%%%%%%%%%%%%%%%%%%%%%%%%%%%%%%%%%%%%%%%%%%%%%%%%%%%%%%%%%%%%%%%%%%%%%%%%%%%%%%%%%%%%%%%%%%%%%%%%%%%%%%%%%%%%%%%%%%%%%%%%%%%%%%%%%%%%%%%%%%%%%%%%%%%%%%%%%%%%%%%%%%%%%%

\section{Discussion}
\label{section:Discussion}

\subsection{Best-fitting mass model}

\begin{figure}
    \centering
    \includegraphics[scale=0.55]{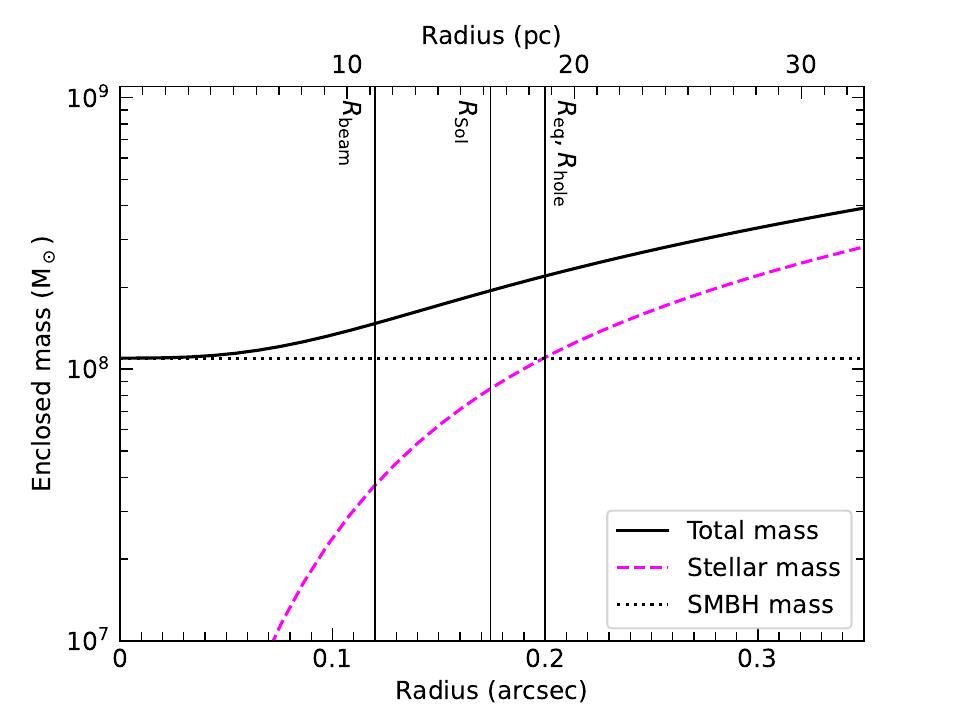}
    \caption{Cumulative mass function of NGC~1387, showing the relative contributions of the SMBH (black dotted line) and stars (magenta dashed line) to the total enclosed mass (black solid line). The three vertical black lines indicate the physical extent of the synthesised beam ($R_\text{beam}$), the usually-defined radius of the SMBH sphere of influence ($R_\text{SoI}$; adopting $\sigma_\text{e}=170.2$~km~s$^{-1}$ and our best-fitting SMBH mass) and the approximate radius of the central hole ($R_\text{hole}$) which is in this case the same as the radius of equal mass contribution ($R_\text{eq}$).}
    \label{fig:cummulative mass function}
\end{figure}

Figure~\ref{fig:cummulative mass function} shows the cumulative mass distribution of NGC~1387. Using the standard approximation of a SMBH SoI (see Section~\ref{section:NGC1387}) and the best-fitting SMBH mass yields $R_\text{SoI}\approx16$~pc ($\approx0.17$~arcsec). One can also assess the impact of a SMBH by considering the radius at which the enclosed stellar mass is equal to that of the SMBH. Figure~\ref{fig:cummulative mass function} shows this radius of equal mass contribution to be $R_\text{eq}\approx19$~pc ($\approx0.20$~arcsec), slightly larger than $R_\text{SoI}$. Whilst our observations of NGC~1387 have a fairly high angular and thus spatial resolution, spatially resolving the SMBH SoI by a factor of two, the smallest scale probed is restricted by the presence of the central hole, which is greater than the size of the synthesised beam. We estimate the radius of this hole by eye to be $\approx0.2$~arcsec ($\approx19$~pc) and have shown it as such in Fig.~\ref{fig:cummulative mass function}. We have chosen not to report the truncation radius obtained from the modelling of the our lower-resolution data cube (see Section~\ref{subsection: inclination prior}), as the synthesised beam of that data cube is larger than the hole and thus does not provide a reliable measure of the hole size. Nevertheless, since the innermost radius probed (the radius of the hole $R_\text{hole}$) is approximately the same as the radius of equal mass contribution, we can conclude that we do in fact measure the effects of the SMBH directly. At this innermost radius, $\approx49$~per cent of the enclosed mass is due to the SMBH.

\begin{figure}
    \centering
    \includegraphics[scale=0.49, trim = {1.6cm 0 0 0}]{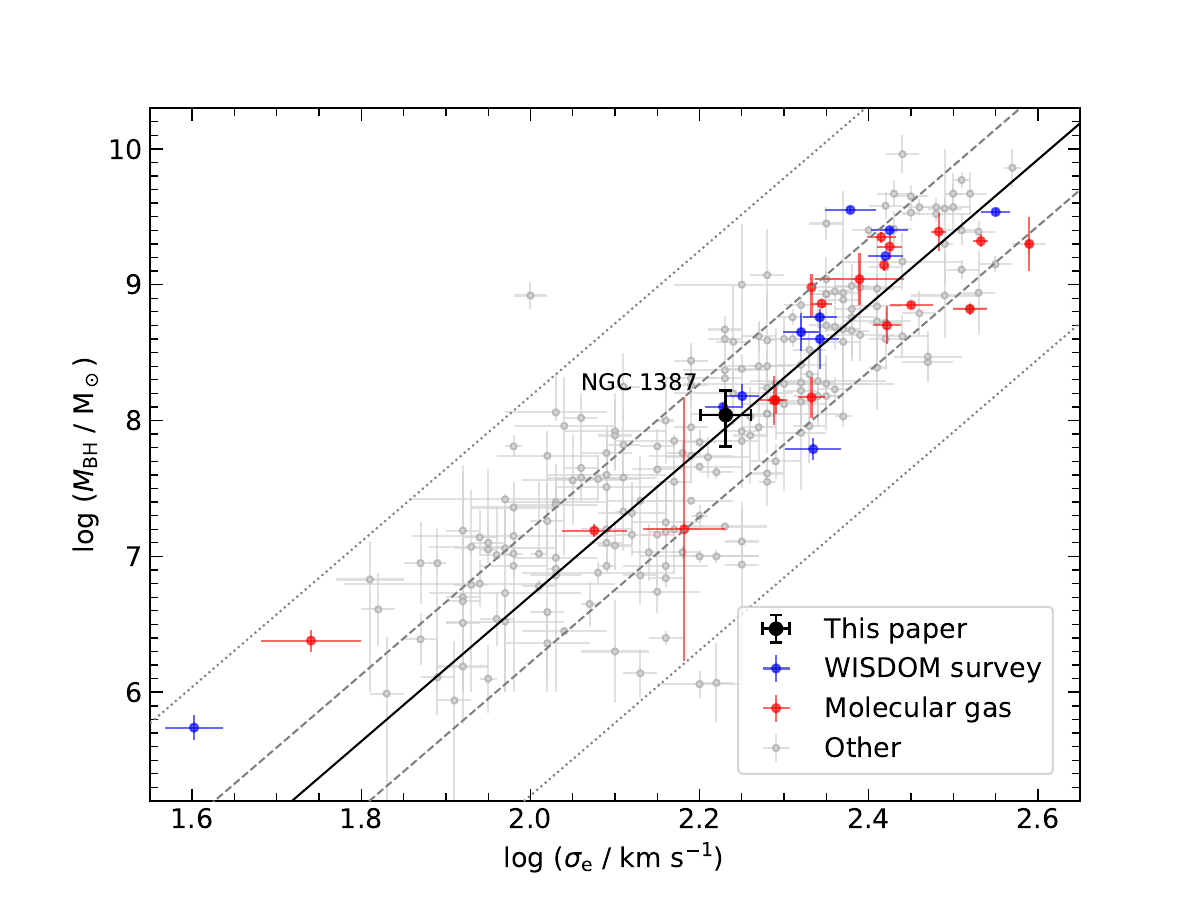}
    \caption{Black hole mass -- stellar velocity dispersion relation obtained using stellar kinematic, ionised-gas kinematic, maser kinematic and active galactic nucleus reverberation mapping measurements (grey data points), as reported by \citet{Bosch_2016}. Blue data points show measurements obtained using molecular gas kinematics by the WISDOM survey, while red data points show measurements obtained using molecular gas kinematics by other groups. The NGC~1387 measurement from this paper is shown as a black data point. Error bars denote $1\sigma$ uncertainties. The solid black line shows the best-fitting relation of \citet{Bosch_2016}, while dashed and dotted grey lines show the $1\sigma$ and $3\sigma$ scatter of the relation, respectively.}
    \label{fig:m-sigma}
\end{figure}

As outlined in Section~\ref{section:Introduction}, accurate determinations of SMBH masses aim to elucidate the processes that govern host galaxy -- SMBH co-evolution. We thus compare our best-fitting NGC~1387 SMBH mass measurement to the $M_{\text{BH}}$ -- $\sigma_\text{e}$ relation of \cite{Bosch_2016} in Fig.~\ref{fig:m-sigma}, revealing that our best-fitting SMBH mass is in very good agreement with the relation.

%%%%%%%%%%%%%%%%%%%%%%%%%%%%%%%%%%%%%%%%%%%%%%%%%%%%%%%%%%%%%%%%%%%%%%%%%%%%%%%%%%%%%%%%%%%%%%%%%%%%
\subsection{Systematic uncertainties}
\label{subsection:systematic uncertainties}

As discussed in Section~\ref{section:Dynamical Modelling}, we have rescaled our statistical uncertainties to approximately account for systematic uncertainties. However, these systematic uncertainties can also be quantified through the creation of models with alternative assumptions and the comparison of those models against our fiducial (i.e.\ best-fitting) model (the one presented in Section~\ref{subsection:results}).

A major source of uncertainty is the dust present in the \textit{HST} image used to create the MGE model. The issue of dust correction has increasingly been considered in the field of molecular-gas SMBH mass determinations \citep[e.g][]{Boizelle2019,Boizelle_2021, Cohn_2024} and in some cases it has been shown to affect the determined SMBH mass by $\approx30$~per cent \citep{Cohn_2021}. Due to the face-on nature of NGC~1387 and its rather uniform dust distribution, we are unable to manually mask the dust. We can instead deal with the effects of dust by performing a dust correction. For this we use a WFC3 F110W-filter ($J$-band) image alongside the F160W-filter ($H$-band) image. We first convert the images to units of magnitude in the Vega system and subtract the F160W-filter image from the F110W-filter image to calculate the observed $J-H$ colour of every pixel. We then examine the outskirts of this image, away from the dustiest regions, to estimate the intrinsic $J-H$ colour of the galaxy, $(J-H)_\text{int}=0.75$. Subtracting this from the observed colour map generates an excess colour map $E(J-H)=(J-H)_\text{obs}-(J-H)_\text{int}$ with a typical colour excess of $0.12$~mag peaking at $0.22$~mag. This is then converted into an $H$-band extinction $A_H$ map by dividing it by the ratio of the total extinctions obtained from the standard Galactic extinction curve with a total-to-selective extinction ratio $R_V=3.1$ \citep{Fitzpatrick_1999}, leading to an average $A_H$ of $0.14$~mag with a peak of $0.26$~mag. This extinction map is then applied to the original F160W-filter image to correct the dust. Lastly, this dust-corrected image is used to generate a new MGE model as in Section~\ref{subsection:stellar potential}, and in turn a new dynamical model. This leads to a best-fitting SMBH mass of $1.91^{+1.64}_{-1.22}\times10^8$~M$_\odot$ and a best-fitting $M/L$ of $0.86_{-0.26}^{+0.28}$~M$_\odot$/L$_{\odot,\text{F160W}}$, representing a $74$~per cent increase of the SMBH mass and a $5$~per cent decrease of the $M/L$ compared to our fiducial model. Despite this $74$~per cent increase, the dust-corrected MGE model best-fitting SMBH mass is still consistent with our fiducial SMBH mass within the $3\sigma$ statistical uncertainties reported in Table~\ref{tab:Best Fit}. \review{In fact, $\chi^2_\text{red} = 1.116$ for the model with the dust-corrected MGE, suggesting that the dust correction only marginally improves the overall fit of our model.}

Another source of potential systematic uncertainty is the assumed axial ratio of the MGE Gaussians. The presence of a bar and the subsequently poor fit of the MGE over it renders us unable to use standard statistical measures like $\chi^2$ when determining the axial ratio. The adopted $q'=0.98$ was determined by eye and as such is somewhat uncertain. To investigate the effect of our choice of fixed axial ratio on the final best-fitting parameters, we generated a series of best-fitting dynamical models whose MGE models had different fixed axial ratios in the range $0.94\leq q'\leq1.00$. The upper boundary of this range simply represents a perfectly circular MGE, whilst the lower boundary represents the lowest $q'$ that still yields a good fit by eye. In all these cases the MGE models were oriented along the same kinematic PA of $242\dotdeg5$. The best-fitting parameters of these models are listed in Table~\ref{tab:systematics}. All the models are statistically consistent with one another and the differences between the SMBH masses and $M/L$ of these models and those of our fiducial model are significantly smaller than the $3\sigma$ statistical uncertainties reported on our fiducial model.

\begin{table}
	\centering
	\caption{Best-fitting parameters and associated uncertainties of models with different fixed MGE axial ratios.}
	\label{tab:systematics}
	\begin{tabular}{ccccc} 
	\hline
	$q'$ & \multicolumn{2}{c}{$\log(M_\text{BH}/\text{M}_\odot)$} & \multicolumn{2}{c}{$M/L_{\text{F160W}}$ (M$_\odot$/L$_{\odot,\text{F160W}}$)}\\
    & Best fit & $3\sigma$ uncertainty & Best fit & $3\sigma$ uncertainty\\
	\hline
        $0.94$ & $8.02$ & $-0.84,+0.37$ & $0.90$ & $-0.36, +0.38$ \\
        $0.96$ & $8.06$ & $-0.85,+0.36$ & $0.86$ & $-0.24, +0.39$ \\
        $0.98$ & $8.04$ & $-0.87,+0.41$ & $0.90$ & $-0.35, +0.44$ \\
        $1.00$ & $8.06$ & $-0.94,+0.43$ & $0.92$ & $-0.33, +0.43$ \\
	\hline
    \end{tabular}
    \begin{tablenotes}
        \item \textit{Note:} The $q'=0.98$ model is the fiducial best-fitting model adopted\\ in this paper. 
    \end{tablenotes}
\end{table}

An additional source of systematic uncertainty stems from our assumptions about the $M/L$ radial profile. Whilst it is typically reasonable to assume that $M/L$ is constant across galaxies, we can nevertheless create models with alternative $M/L$ profiles to assess the effects this might have on the SMBH mass. In particular, there is a bright central ring of ionised-gas emission in NGC~1387, suggesting star-forming regions \citep{Iodice_2019}. We thus tested a linearly-varying $M/L$ radial profile. We implemented this by creating a circular velocity curve assuming $M/L=1$~M$_\odot/\text{L}_{\odot,\text{F160W}}$, and then scaling these velocities by the square root of the $M/L$ profile. Keeping all other parameters as before, a model with a radially linearly-varying $M/L$ yields a best-fitting SMBH mass of $4.3^{+11.5}_{-4.2}\times10^7$~M$_\odot$, a $61$~per cent decrease compared to out fiducial model, but nevertheless consistent with it within the $3\sigma$ statistical uncertainties reported in Table~\ref{tab:Best Fit}. The best-fitting $M/L$ is $\left(M/L_\text{F160W}\right)/\left(\text{M}_\odot/\text{L}_{\odot,\text{F160W}}\right)=0.83^{+0.28}_{-0.22}-0.02^{+0.02}_{-0.02}\,\left(R/\text{arcsec}\right)$, a decrease of $8$~per cent in the very centre. Whilst this result is consistent with our fiducial model within even $1\sigma$ statistical uncertainties, it is somewhat unexpected, as due to conservation of total dynamical mass a smaller SMBH would imply a larger $M/L$ whereas this $M/L$ is smaller than that of our fiducial model. However, this result can be explained by the best-fit inclination of $15.6\degree^{+2.7}_{-2.2}$, an increase of $10$~per~cent compared to our fiducial model. Looking at Fig.~\ref{fig:cornerPlot} an inclination change of $10\%$ should lead to a change in $M/L$ consistent with the above result. \review{The model with a linearly-varying $M/L$ has $\chi^2_\text{red}=1.108$, which is an improvement compared to the model with a constant $M/L$ and the model with a dust-corrected MGE. This could suggest that the dust correction does not adequately correct the innermost parts of the surface brightness distribution, and a linearly-varying $M/L$ better captures the true stellar mass surface density profile of the galaxy.}

\review{Taking these tests together, it is clear that the systematic variations of the SMBH mass are similar to, albeit slightly larger than, the statistical uncertainties reported in Table~\ref{tab:Best Fit}, with $1.10^{+1.71}_{-0.95}[\text{stat},3\sigma]^{+2.45}_{-1.09}[\text{sys}]\times10^8$~M$_\odot$, while the systematic variations of the $M/L$ are in line with the statistical uncertainties, with $0.90^{+0.44}_{-0.35}[\text{stat}, 3\sigma]^{+0.46}_{-0.36}[\text{sys}]$~M$_\odot$/L$_{\odot,\text{F160W}}$. For each parameter we adopt the maximum upper bound and minimum lower bound obtained among all model variants, therefore representing the largest deviations from the fiducial model permitted by the explored systematic tests.}

Aside from direct assumptions we make in our model, an important source of systematic error is the adopted galaxy distance. This is largely dependent on the distance-measurement method used, with numerical action methods presenting discrepancies of a factor of $2$ \citep{McQuinn_2021}, whilst Cepheid and tip of the red giant branch distance indicators can routinely yield uncertainties of $<5$~per cent \citep{Riess_2024}. The distance used in this paper was determined using surface brightness fluctuations, which typically yield distances accurate to $\approx5$~per cent \citep{Cantiello_2023}, consistent with the stated uncertainties on the NGC~1387 distance measurement. As is standard practice, however, we do not include the distance uncertainty in our final uncertainty budget, as our fiducial mass can be trivially (linearly) scaled to any other adopted distance.

%%%%%%%%%%%%%%%%%%%%%%%%%%%%%%%%%%%%%%%%%%%%%%%%%%%%%%%%%%%%%%%%%%%%%%%%%%%%%%%%%%%%%%%%%%%%%%%%%%%%

%\subsection{The $M_{\text{BH}}-\sigma$ relation}

%%%%%%%%%%%%%%%%%%%%%%%%%%%%%%%%%%%%%%%%%%%%%%%%%%%%%%%%%%%%%%%%%%%%%%%%%%%%%%%%%%%%%%%%%%%%%%%%%%%%

\section{Conclusions}
\label{section:Conclusion}

High angular resolution ALMA $^{12}$CO(2–1) observations of the ETG NGC~1387 were obtained, mapping its nearly face-on regularly-rotating molecular gas disc. The stellar mass distribution was estimated using a MGE model of a \textit{HST} image and a spatially-constant $M/L$. The molecular gas distribution and kinematics were then forward modelled using \textsc{KinMS} and a MCMC framework, yielding a best-fitting SMBH mass of \review{$1.10^{+1.71}_{-0.95}[\text{stat},3\sigma]^{+2.45}_{-1.09}[\text{sys}]\times10^8$~M$_\odot$} and a stellar F160W-filter $M/L$ of \review{$0.90^{+0.44}_{-0.35}[\text{stat}, 3\sigma]^{+0.46}_{-0.36}[\text{sys}]$~M$_\odot$/L$_{\odot,\text{F160W}}$}. This inferred SMBH mass is in good agreement with the $M_\text{BH}$ -- $\sigma_\text{e}$ relation of \citet{Bosch_2016}. 

%%%%%%%%%%%%%%%%%%%%%%%%%%%%%%%%%%%%%%%%%%%%%%%%%%%%%%%%%%%%%%%%%%%%%%%%%%%%%%%%%%%%%%%%%%%%%%%%%%%%
\section*{Acknowledgements}

PD acknowledges support from a Science and Technology Facilities Council (STFC) DPhil studentship under grant ST/S505638/1. MB was supported by STFC consolidated grant “Astrophysics at Oxford” ST/K00106X/1 and  ST/W000903/1. TAD acknowledges support from STFC through grants ST/S00033X/1 and ST/W000830/1. This paper makes use of the following ALMA data: ADS/JAO.ALMA 2016.1.00437.S and 2016.2.00053.S. ALMA is a partnership of ESO (representing its member states), NSF (USA), and NINS (Japan), together with NRC (Canada), MOST and ASIAA (Taiwan), and KASI (Republic of Korea), in cooperation with the Republic of Chile. The Joint ALMA Observatory is operated by ESO, AUI/NRAO, and NAOJ. This research made use of the NASA/IPAC Extragalactic Database (NED), which is operated by the Jet Propulsion Laboratory, California Institute of Technology, under contract with the National Aeronautics and Space Administration.

%%%%%%%%%%%%%%%%%%%%%%%%%%%%%%%%%%%%%%%%%%%%%%%%%%%%%%%%%%%%%%%%%%%%%%%%%%%%%%%%%%%%%%%%%%%%%%%%%%%%
%%%%%%%%%%%%%%%%%%%%%%%%%%%%%%%%%%%%%%%%%%%%%%%%%%%%%%%%%%%%%%%%%%%%%%%%%%%%%%%%%%%%%%%%%%%%%%%%%%%%

\section*{Data Availability}
 
The observations underlying this article are available in the ALMA archive, at \url{https://almascience.eso.org/asax/}, and in the Hubble Science Archive, at \url{https://hst.esac.esa.int/ehst/}.

%%%%%%%%%%%%%%%%%%%% REFERENCES %%%%%%%%%%%%%%%%%%%%%%%%%%%%%%%%%%%%%%%%%%%%%%%%%%%%%%%%%%%%%%%%%%%%

\bibliographystyle{mnras}
\bibliography{BIB} % if your bibtex file is called example.bib

@INPROCEEDINGS{McMullin,
       author = {{McMullin}, J.~P. and {Waters}, B. and {Schiebel}, D. and {Young}, W. and {Golap}, K.},
        title = "{CASA Architecture and Applications}",
    booktitle = {Astronomical Data Analysis Software and Systems XVI},
         year = 2007,
       editor = {{Shaw}, R.~A. and {Hill}, F. and {Bell}, D.~J.},
       series = {Astronomical Society of the Pacific Conference Series},
       volume = {376},
        month = oct,
        pages = {127},
       url = {https://ui.adsabs.harvard.edu/abs/2007ASPC..376..127M}
}

@article{Davis2013a,
    author = {Davis, Timothy A. and Alatalo, Katherine and Bureau, Martin and Cappellari, Michele and Scott, Nicholas and Young, Lisa M. and Blitz, Leo and Crocker, Alison and Bayet, Estelle and Bois, Maxime and Bournaud, Frédéric and Davies, Roger L. and de Zeeuw, P. T. and Duc, Pierre-Alain and Emsellem, Eric and Khochfar, Sadegh and Krajnović, Davor and Kuntschner, Harald and Lablanche, Pierre-Yves and McDermid, Richard M. and Morganti, Raffaella and Naab, Thorsten and Oosterloo, Tom and Sarzi, Marc and Serra, Paolo and Weijmans, Anne-Marie},
    title = "{The ATLAS3D Project – XIV. The extent and kinematics of the molecular gas in early-type galaxies}",
    journal = {MNRAS},
    volume = {429},
    number = {1},
    pages = {534-555},
    year = {2013},
    month = {12}, 
    doi = {10.1093/mnras/sts353}
}

@article{Davis2017,
    author = {Davis, Timothy A. and Bureau, Martin and Onishi, Kyoko and Cappellari, Michele and Iguchi, Satoru and Sarzi, Marc},
    title = "{WISDOM Project – II. Molecular gas measurement of the supermassive black hole mass in NGC 4697}",
    journal = {MNRAS},
    volume = {468},
    number = {4},
    pages = {4675-4690},
    year = {2017},
    month = {04}, 
    doi = {10.1093/mnras/stw3217}
}

@ARTICLE{Cappellari2002,
       author = {{Cappellari}, Michele},
        title = "{Efficient multi-Gaussian expansion of galaxies}",
      journal = {\mnras},
         year = 2002,
        month = jun,
       volume = {333},
       number = {2},
        pages = {400-410},
          doi = {10.1046/j.1365-8711.2002.05412.x},
archivePrefix = {arXiv},
       eprint = {astro-ph/0201430},
 primaryClass = {astro-ph},
       url = {https://ui.adsabs.harvard.edu/abs/2002MNRAS.333..400C}
}

@article{Emsellem1994,
       author = {{Emsellem}, E. and {Monnet}, G. and {Bacon}, R.},
        title = "{The multi-gaussian expansion method: a tool for building realistic photometric and kinematical models of stellar systems I. The formalism}",
      journal = {\aap},
         year = 1994,
        month = may,
       volume = {285},
        pages = {723-738}, 
        url = {https://ui.adsabs.harvard.edu/abs/1994A&A...285..723E}
}

@INPROCEEDINGS{TinyTim,
       author = {{Krist}, John E. and {Hook}, Richard N. and {Stoehr}, Felix},
        title = "{20 years of Hubble Space Telescope optical modeling using Tiny Tim}",
    booktitle = {Optical Modeling and Performance Predictions V},
         year = 2011,
       editor = {{Kahan}, Mark A.},
       series = {Society of Photo-Optical Instrumentation Engineers (SPIE) Conference Series},
       volume = {8127},
        month = oct,
          eid = {81270J},
        pages = {81270J}
}

@article{Willmer2018,
	year = 2018,
	month = {jun},
	publisher = {American Astronomical Society},
	volume = {236},
	number = {2},
	pages = {47},
	author = {Christopher N. A. Willmer},
	title = {The Absolute Magnitude of the Sun in Several Filters},
	journal = {ApJSS},
    doi = {10.3847/1538-4365/aabfdf}
}

@ARTICLE{Kormendy2013,
       author = {{Kormendy}, John and {Ho}, Luis C.},
        title = "{Coevolution (Or Not) of Supermassive Black Holes and Host Galaxies}",
      journal = {\araa},
         year = 2013,
        month = aug,
       volume = {51},
       number = {1},
        pages = {511-653},
          doi = {10.1146/annurev-astro-082708-101811},
archivePrefix = {arXiv},
       eprint = {1304.7762},
 primaryClass = {astro-ph.CO},
       url = {https://ui.adsabs.harvard.edu/abs/2013ARA&A..51..511K}
}

@ARTICLE{Kuo2011,
       author = {{Kuo}, C.~Y. and {Braatz}, J.~A. and {Condon}, J.~J. and {Impellizzeri}, C.~M.~V. and {Lo}, K.~Y. and {Zaw}, I. and {Schenker}, M. and {Henkel}, C. and {Reid}, M.~J. and {Greene}, J.~E.},
        title = "{The Megamaser Cosmology Project. III. Accurate Masses of Seven Supermassive Black Holes in Active Galaxies with Circumnuclear Megamaser Disks}",
      journal = {\apj},
         year = 2011,
        month = jan,
       volume = {727},
       number = {1},
          eid = {20},
        pages = {20},
          doi = {10.1088/0004-637X/727/1/20},
archivePrefix = {arXiv},
       eprint = {1008.2146},
 primaryClass = {astro-ph.CO}
}

@ARTICLE{Gao2017,
       author = {{Gao}, F. and {Braatz}, J.~A. and {Reid}, M.~J. and {Condon}, J.~J. and {Greene}, J.~E. and {Henkel}, C. and {Impellizzeri}, C.~M.~V. and {Lo}, K.~Y. and {Kuo}, C.~Y. and {Pesce}, D.~W. and {Wagner}, J. and {Zhao}, W.},
        title = "{The Megamaser Cosmology Project. IX. Black Hole Masses for Three Maser Galaxies}",
      journal = {\apj},
         year = 2017,
        month = jan,
       volume = {834},
       number = {1},
          eid = {52},
        pages = {52},
          doi = {10.3847/1538-4357/834/1/52},
archivePrefix = {arXiv},
       eprint = {1610.06802},
 primaryClass = {astro-ph.GA},
       url = {https://ui.adsabs.harvard.edu/abs/2017ApJ...834...52G}
}

@ARTICLE{Krajnovic2009,
       author = {{Krajnovi{\'c}}, Davor and {McDermid}, Richard M. and {Cappellari}, Michele and {Davies}, Roger L.},
        title = "{Determination of masses of the central black holes in NGC 524 and 2549 using laser guide star adaptive optics}",
      journal = {\mnras},
         year = 2009,
        month = nov,
       volume = {399},
       number = {4},
        pages = {1839-1857},
          doi = {10.1111/j.1365-2966.2009.15415.x},
archivePrefix = {arXiv},
       eprint = {0907.3748},
 primaryClass = {astro-ph.GA},
       url = {https://ui.adsabs.harvard.edu/abs/2009MNRAS.399.1839K}
}

@ARTICLE{Barth2016,
       author = {{Barth}, Aaron J. and {Boizelle}, Benjamin D. and {Darling}, Jeremy and {Baker}, Andrew J. and {Buote}, David A. and {Ho}, Luis C. and {Walsh}, Jonelle L.},
        title = "{Measurement of the Black Hole Mass in NGC 1332 from ALMA Observations at 0.044 arcsecond Resolution}",
      journal = {ApJ},
         year = 2016,
        month = may,
       volume = {822},
       number = {2},
          eid = {L28},
        pages = {L28},
          doi = {10.3847/2041-8205/822/2/L28},
archivePrefix = {arXiv},
       eprint = {1605.01346},
 primaryClass = {astro-ph.GA},
       url = {https://ui.adsabs.harvard.edu/abs/2016ApJ...822L..28B}
}

@ARTICLE{Nagai2019,
       author = {{Nagai}, H. and {Onishi}, K. and {Kawakatu}, N. and {Fujita}, Y. and {Kino}, M. and {Fukazawa}, Y. and {Lim}, J. and {Forman}, W. and {Vrtilek}, J. and {Nakanishi}, K. and {Noda}, H. and {Asada}, K. and {Wajima}, K. and {Ohyama}, Y. and {David}, L. and {Daikuhara}, K.},
        title = "{The ALMA Discovery of the Rotating Disk and Fast Outflow of Cold Molecular Gas in NGC 1275}",
      journal = {\apj},
         year = 2019,
        month = oct,
       volume = {883},
       number = {2},
          eid = {193},
        pages = {193},
          doi = {10.3847/1538-4357/ab3e6e},
archivePrefix = {arXiv},
       eprint = {1905.06017},
 primaryClass = {astro-ph.GA},
       url = {https://ui.adsabs.harvard.edu/abs/2019ApJ...883..193N}
}

@ARTICLE{Boizelle2019,
       author = {{Boizelle}, Benjamin D. and {Barth}, Aaron J. and {Walsh}, Jonelle L. and {Buote}, David A. and {Baker}, Andrew J. and {Darling}, Jeremy and {Ho}, Luis C.},
        title = "{A Precision Measurement of the Mass of the Black Hole in NGC 3258 from High-resolution ALMA Observations of Its Circumnuclear Disk}",
      journal = {\apj},
         year = 2019,
        month = aug,
       volume = {881},
       number = {1},
          eid = {10},
        pages = {10},
          doi = {10.3847/1538-4357/ab2a0a},
archivePrefix = {arXiv},
       eprint = {1906.06267},
 primaryClass = {astro-ph.GA},
       url = {https://ui.adsabs.harvard.edu/abs/2019ApJ...881...10B}
}

@article{WISDOM_V,
    author = {North, Eve V and Davis, Timothy A and Bureau, Martin and Cappellari, Michele and Iguchi, Satoru and Liu, Lijie and Onishi, Kyoko and Sarzi, Marc and Smith, Mark D and Williams, Thomas G},
    title = "{WISDOM project – V. Resolving molecular gas in Keplerian rotation around the supermassive black hole in NGC 0383}",
    journal = {MNRAS},
    volume = {490},
    number = {1},
    pages = {319-330},
    year = {2019},
    month = {09},
    issn = {0035-8711},
    doi = {10.1093/mnras/stz2598},
    url = {https://doi.org/10.1093/mnras/stz2598},
    eprint = {https://academic.oup.com/mnras/article-pdf/490/1/319/30100460/stz2598.pdf},
}

@article{WISDOM_IV,
    author = {Smith, Mark D and Bureau, Martin and Davis, Timothy A and Cappellari, Michele and Liu, Lijie and North, Eve V and Onishi, Kyoko and Iguchi, Satoru and Sarzi, Marc},
    title = "{WISDOM project – IV. A molecular gas dynamical measurement of the supermassive black hole mass in NGC 524}",
    journal = {MNRAS},
    volume = {485},
    number = {3},
    pages = {4359-4374},
    year = {2019},
    month = {03},
    issn = {0035-8711},
    doi = {10.1093/mnras/stz625},
    url = {https://doi.org/10.1093/mnras/stz625},
    eprint = {https://academic.oup.com/mnras/article-pdf/485/3/4359/28231428/stz625.pdf},
}

@article{Bosch_2016,
	doi = {10.3847/0004-637x/831/2/134},
	url = {https://doi.org/10.3847/0004-637x/831/2/134},
	year = 2016,
	month = {nov},
	publisher = {American Astronomical Society},
	volume = {831},
	number = {2},
	pages = {134},
	author = {Remco C. E. van den Bosch},
	title = {{UNIFICATION} {OF} {THE} {FUNDAMENTAL} {PLANE} {AND} {SUPER} {MASSIVE} {BLACK} {HOLE} {MASSES}},
	journal = {ApJ}
}

@misc{Dame_2011,
  doi = {10.48550/ARXIV.1101.1499},
  
  url = {https://arxiv.org/abs/1101.1499},
  
  author = {Dame, T. M.},
  
  title = {Optimization of Moment Masking for CO Spectral Line Surveys},
  
  publisher = {arXiv},
  
  year = {2011},
  
  copyright = {arXiv.org perpetual, non-exclusive license}
}

@misc{WFC3, 
url = {https://hst-docs.stsci.edu/wfc3dhb},
author = {Sahu, K. C. and Anderson, J. and Baggett, S.}, 
year = {2021}, 
title = {WFC3 Data Handbook, v5.0}, 
publisher = {Baltimore MD: STScI}
}

@article{Davis_2013b,
	doi = {10.1038/nature11819},
  
	url = {https://doi.org/10.1038%2Fnature11819},
  
	year = 2013,
	month = {jan},
  
	publisher = {Springer Science and Business Media {LLC}
},
  
	volume = {494},
  
	number = {7437},
  
	pages = {328--330},
  
	author = {Timothy A. Davis and Martin Bureau and Michele Cappellari and Marc Sarzi and Leo Blitz},
  
	title = {A black-hole mass measurement from molecular gas kinematics in {NGC}4526},
  
	journal = {Nature}
}

@article{Onishi_2015,
	doi = {10.1088/0004-637x/806/1/39},
  
	url = {https://doi.org/10.1088%2F0004-637x%2F806%2F1%2F39},
  
	year = 2015,
	month = {jun},
  
	publisher = {American Astronomical Society},
  
	volume = {806},
  
	number = {1},
  
	pages = {39},
  
	author = {K. Onishi and S. Iguchi and K. Sheth and K. Kohno},
  
	title = {A {MEASUREMENT} {OF} {THE} {BLACK} {HOLE} {MASS} {IN} {NGC} 1097 {USING} {ALMA}
},
  
	journal = {ApJ}
}

@article{WISDOM_I,
	doi = {10.1093/mnras/stx631},
  
	url = {https://doi.org/10.1093%2Fmnras%2Fstx631},
  
	year = 2017,
	month = {mar},
  
	publisher = {Oxford University Press ({OUP})},
  
	volume = {468},
  
	number = {4},
  
	pages = {4663--4674},
  
	author = {Kyoko Onishi and Satoru Iguchi and Timothy A. Davis and Martin Bureau and Michele Cappellari and Marc Sarzi and Leo Blitz},
  
	title = {{WISDOM} project {\textendash} I. Black hole mass measurement using molecular gas kinematics in {NGC} 3665},
  
	journal = {MNRAS}
}

@article{WISDOM_III,
	doi = {10.1093/mnras/stx2600},
  
	url = {https://doi.org/10.1093%2Fmnras%2Fstx2600},
  
	year = 2017,
	month = {oct},
  
	publisher = {Oxford University Press ({OUP})},
  
	volume = {473},
  
	number = {3},
  
	pages = {3818--3834},
  
	author = {Timothy A. Davis and Martin Bureau and Kyoko Onishi and Freeke van de Voort and Michele Cappellari and Satoru Iguchi and Lijie Liu and Eve V. North and Marc Sarzi and Mark D. Smith},
  
	title = {{WISDOM} Project {\textendash} {III}. Molecular gas measurement of the supermassive black hole mass in the barred lenticular galaxy {NGC}4429},
  
	journal = {MNRAS}
}

@article{Ruffa_2019,
	doi = {10.1093/mnras/stz2368},
  
	url = {https://doi.org/10.1093%2Fmnras%2Fstz2368},
  
	year = 2019,
	month = {aug},
  
	publisher = {Oxford University Press ({OUP})},
  
	volume = {489},
  
	number = {3},
  
	pages = {3739--3757},
  
	author = {Ilaria Ruffa and Timothy A Davis and Isabella Prandoni and Robert A Laing and Rosita Paladino and Paola Parma and Hans de~Ruiter and Viviana Casasola and Martin Bureau and Joshua Warren},
  
	title = {The {AGN} fuelling/feedback cycle in nearby radio galaxies {\textendash} {II}. Kinematics of the molecular gas},
  
	journal = {MNRAS}
}

@article{Nguyen_2020,
	doi = {10.3847/1538-4357/ab77aa},
  
	url = {https://doi.org/10.3847%2F1538-4357%2Fab77aa},
  
	year = 2020,
	month = {mar},
  
	publisher = {American Astronomical Society},
  
	volume = {892},
  
	number = {1},
  
	pages = {68},
  
	author = {Dieu D. Nguyen and Mark den Brok and Anil C. Seth and Timothy A. Davis and Jenny E. Greene and Michelle Cappellari and Joseph B. Jensen and Sabine Thater and Satoru Iguchi and Masatoshi Imanishi and Takuma Izumi and Kristina Nyland and Nadine Neumayer and Kouichiro Nakanishi and Phuong M. Nguyen and Takafumi Tsukui and Martin Bureau and Kyoko Onishi and Quang L. Nguyen and Ngan M. Le},
  
	title = {The {MBHBM}$\less$sub$\greater$$\star$$\less$/sub$\greater$ Project. I. Measurement of the Central Black Hole Mass in Spiral Galaxy {NGC} 3504 Using Molecular Gas Kinematics},
  
	journal = {ApJ}
}

@article{Smith_2021,
	doi = {10.1093/mnras/stab791},
  
	url = {https://doi.org/10.1093%2Fmnras%2Fstab791},
  
	year = 2021,
	month = {mar},
  
	publisher = {Oxford University Press ({OUP})},
  
	volume = {503},
  
	number = {4},
  
	pages = {5984--5996},
  
	author = {Mark D Smith and Martin Bureau and Timothy A Davis and Michele Cappellari and Lijie Liu and Kyoko Onishi and Satoru Iguchi and Eve V North and Marc Sarzi and Thomas G Williams},
  
	title = {{WISDOM} project {\textendash} {VII}. Molecular gas measurement of the supermassive black hole mass in the elliptical galaxy {NGC} 7052},
  
	journal = {MNRAS}
}

@ARTICLE{Boizelle_2021,
       author = {{Boizelle}, Benjamin D. and {Walsh}, Jonelle L. and {Barth}, Aaron J. and {Buote}, David A. and {Baker}, Andrew J. and {Darling}, Jeremy and {Ho}, Luis C. and {Cohn}, Jonathan and {Kabasares}, Kyle M.},
        title = "{Black Hole Mass Measurements of Radio Galaxies NGC 315 and NGC 4261 Using ALMA CO Observations}",
      journal = {\apj},
         year = 2021,
        month = feb,
       volume = {908},
       number = {1},
          eid = {19},
        pages = {19},
          doi = {10.3847/1538-4357/abd24d},
archivePrefix = {arXiv},
       eprint = {2012.04669},
 primaryClass = {astro-ph.GA},
       url = {https://ui.adsabs.harvard.edu/abs/2021ApJ...908...19B}
}

@article{Davis_2020,
	doi = {10.1093/mnras/staa1567},
  
	url = {https://doi.org/10.1093%2Fmnras%2Fstaa1567},
  
	year = 2020,
	month = {jul},
  
	publisher = {Oxford University Press ({OUP})},
  
	volume = {496},
  
	number = {4},
  
	pages = {4061--4078},
  
	author = {Timothy A Davis and Dieu D Nguyen and Anil C Seth and Jenny E Greene and Kristina Nyland and Aaron J Barth and Martin Bureau and Michele Cappellari and Mark den~Brok and Satoru Iguchi and Federico Lelli and Lijie Liu and Nadine Neumayer and Eve V North and Kyoko Onishi and Marc Sarzi and Mark D Smith and Thomas G Williams},
  
	title = {Revealing the intermediate-mass black hole at the heart of the dwarf galaxy {NGC}{\hspace{0.167em}
}404 with sub-parsec resolution {ALMA} observations},
  
	journal = {MNRAS}
}

@article{Kabasares_2022,
	doi = {10.3847/1538-4357/ac7a38},
  
	url = {https://doi.org/10.3847%2F1538-4357%2Fac7a38},
  
	year = 2022,
	month = {aug},
  
	publisher = {American Astronomical Society},
  
	volume = {934},
  
	number = {2},
  
	pages = {162},
  
	author = {Kyle M. Kabasares and Aaron J. Barth and David A. Buote and Benjamin D. Boizelle and Jonelle L. Walsh and Andrew J. Baker and Jeremy Darling and Luis C. Ho and Jonathan Cohn},
  
	title = {Black Hole Mass Measurements of Early-type Galaxies {NGC} 1380 and {NGC} 6861 through {ALMA} and {HST} Observations and Gas-dynamical Modeling{\ast}
},
  
	journal = {ApJ}
}

@Article{Mitzkus2017,
  author        = {Mitzkus, M. and Cappellari, M. and Walcher, C. J.},
  title         = {{Dominant dark matter and a counter-rotating disc: MUSE view of the low-luminosity S0 galaxy NGC 5102}},
  doi           = {10.1093/mnras/stw2677},
  eprint        = {1610.04516},
  pages         = {4789--4806},
  volume        = {464},
  archiveprefix = {arXiv},
  journal       = {\mnras},
  month         = feb,
  year          = {2017},
}

@ARTICLE{McConnell2012,
       author = {{McConnell}, Nicholas J. and {Ma}, Chung-Pei and {Murphy}, Jeremy D. and {Gebhardt}, Karl and {Lauer}, Tod R. and {Graham}, James R. and {Wright}, Shelley A. and {Richstone}, Douglas O.},
        title = "{Dynamical Measurements of Black Hole Masses in Four Brightest Cluster Galaxies at 100 Mpc}",
      journal = {\apj},
         year = 2012,
        month = sep,
       volume = {756},
       number = {2},
          eid = {179},
        pages = {179},
          doi = {10.1088/0004-637X/756/2/179},
archivePrefix = {arXiv},
       eprint = {1203.1620},
 primaryClass = {astro-ph.CO},
       url = {https://ui.adsabs.harvard.edu/abs/2012ApJ...756..179M}
}

@ARTICLE{Nguyen_2022,
       author = {{Nguyen}, Dieu D. and {Bureau}, Martin and {Thater}, Sabine and {Nyland}, Kristina and {den Brok}, Mark and {Cappellari}, Michele and {Davis}, Timothy A. and {Greene}, Jenny E. and {Neumayer}, Nadine and {Imanishi}, Masatoshi and {Izumi}, Takuma and {Kawamuro}, Taiki and {Baba}, Shunsuke and {Nguyen}, Phuong M. and {Iguchi}, Satoru and {Tsukui}, Takafumi and {Lam}, T.~N. and {Ho}, Than},
        title = "{The MBHBM$^{{\ensuremath{\star}}}$ Project - II. Molecular gas kinematics in the lenticular galaxy NGC 3593 reveal a supermassive black hole}",
      journal = {MNRAS},
         year = 2022,
        month = jan,
       volume = {509},
       number = {2},
        pages = {2920-2939},
          doi = {10.1093/mnras/stab3016},
archivePrefix = {arXiv},
       eprint = {2110.08476},
 primaryClass = {astro-ph.GA}
}

@ARTICLE{Nguyen_2021,
       author = {{Nguyen}, Dieu D. and {Izumi}, Takuma and {Thater}, Sabine and {Imanishi}, Masatoshi and {Kawamuro}, Taiki and {Baba}, Shunsuke and {Nakano}, Suzuka and {Turner}, Jean L. and {Kohno}, Kotaro and {Matsushita}, Satoki and {Mart{\'\i}n}, Sergio and {Meier}, David S. and {Nguyen}, Phuong M. and {Nguyen}, Lam T.},
        title = "{Black hole mass measurement using ALMA observations of [CI] and CO emissions in the Seyfert 1 galaxy NGC 7469}",
      journal = {MNRAS},
         year = 2021,
        month = jul,
       volume = {504},
       number = {3},
        pages = {4123-4142},
          doi = {10.1093/mnras/stab1002},
archivePrefix = {arXiv},
       eprint = {2104.03539},
 primaryClass = {astro-ph.GA}
}

@article{Ruffa_2023,
	doi = {10.1093/mnras/stad1119},
  
	url = {https://doi.org/10.1093%2Fmnras%2Fstad1119},
  
	year = 2023,
	month = {may},
  
	publisher = {Oxford University Press ({OUP})},
  
	volume = {522},
  
	number = {4},
  
	pages = {6170--6195},
  
	author = {Ilaria Ruffa and Timothy A Davis and Michele Cappellari and Martin Bureau and Jacob Elford and Satoru Iguchi and Federico Lelli and Fu-Heng Liang and Lijie Liu and Anan Lu and Marc Sarzi and Thomas G Williams},
  
	title = {{WISDOM} project {\textendash} {XIV}. {SMBH} mass in the early-type galaxies {NGC}{\hspace{0.167em}
}0612, {NGC}{\hspace{0.167em}}1574, and {NGC}{\hspace{0.167em}}4261 from {CO} dynamical modelling},
  
	journal = {MNRAS}
}

@ARTICLE{Schlafly_2011,
       author = {{Schlafly}, Edward F. and {Finkbeiner}, Douglas P.},
        title = "{Measuring Reddening with Sloan Digital Sky Survey Stellar Spectra and Recalibrating SFD}",
      journal = {\apj},
         year = 2011,
        month = aug,
       volume = {737},
       number = {2},
          eid = {103},
        pages = {103},
          doi = {10.1088/0004-637X/737/2/103},
archivePrefix = {arXiv},
       eprint = {1012.4804},
 primaryClass = {astro-ph.GA}
}

@article{Cohn_2021,
doi = {10.3847/1538-4357/ac0f78},
url = {https://dx.doi.org/10.3847/1538-4357/ac0f78},
year = {2021},
month = {sep},
publisher = {The American Astronomical Society},
volume = {919},
number = {2},
pages = {77},
author = {Jonathan H. Cohn and Jonelle L. Walsh and Benjamin D. Boizelle and Aaron J. Barth and Karl Gebhardt and Kayhan Gültekin and Akın Yıldırım and David A. Buote and Jeremy Darling and Andrew J. Baker and Luis C. Ho and Kyle M. Kabasares},
title = {An ALMA Gas-dynamical Mass Measurement of the Supermassive Black Hole in the Local Compact Galaxy UGC 2698},
journal = {ApJ}

}

@ARTICLE{McConnell_2013,
       author = {{McConnell}, Nicholas J. and {Ma}, Chung-Pei},
        title = "{Revisiting the Scaling Relations of Black Hole Masses and Host Galaxy Properties}",
      journal = {\apj},
         year = 2013,
        month = feb,
       volume = {764},
       number = {2},
          eid = {184},
        pages = {184},
          doi = {10.1088/0004-637X/764/2/184},
archivePrefix = {arXiv},
       eprint = {1211.2816},
 primaryClass = {astro-ph.CO},
       adsurl = {https://ui.adsabs.harvard.edu/abs/2013ApJ...764..184M}
}

@ARTICLE{Gebhardt_2000,
       author = {{Gebhardt}, Karl and {Bender}, Ralf and {Bower}, Gary and {Dressler}, Alan and {Faber}, S.~M. and {Filippenko}, Alexei V. and {Green}, Richard and {Grillmair}, Carl and {Ho}, Luis C. and {Kormendy}, John and {Lauer}, Tod R. and {Magorrian}, John and {Pinkney}, Jason and {Richstone}, Douglas and {Tremaine}, Scott},
        title = "{A Relationship between Nuclear Black Hole Mass and Galaxy Velocity Dispersion}",
      journal = {\apjl},
         year = 2000,
        month = aug,
       volume = {539},
       number = {1},
        pages = {L13-L16},
          doi = {10.1086/312840},
archivePrefix = {arXiv},
       eprint = {astro-ph/0006289},
 primaryClass = {astro-ph},
       adsurl = {https://ui.adsabs.harvard.edu/abs/2000ApJ...539L..13G}
}

@ARTICLE{Blakeslee_2009,
       author = {{Blakeslee}, John P. and {Jord{\'a}n}, Andr{\'e}s and {Mei}, Simona and {C{\^o}t{\'e}}, Patrick and {Ferrarese}, Laura and {Infante}, Leopoldo and {Peng}, Eric W. and {Tonry}, John L. and {West}, Michael J.},
        title = "{The ACS Fornax Cluster Survey. V. Measurement and Recalibration of Surface Brightness Fluctuations and a Precise Value of the Fornax-Virgo Relative Distance}",
      journal = {\apj},
         year = 2009,
        month = mar,
       volume = {694},
       number = {1},
        pages = {556-572},
          doi = {10.1088/0004-637X/694/1/556},
archivePrefix = {arXiv},
       eprint = {0901.1138},
 primaryClass = {astro-ph.CO}
}

@ARTICLE{Wegner_2003,
       author = {{Wegner}, G. and {Bernardi}, M. and {Willmer}, C.~N.~A. and {da Costa}, L.~N. and {Alonso}, M.~V. and {Pellegrini}, P.~S. and {Maia}, M.~A.~G. and {Chaves}, O.~L. and {Rit{\'e}}, C.},
        title = "{Redshift-Distance Survey of Early-Type Galaxies: Spectroscopic Data}",
      journal = {AJ},
         year = 2003,
        month = nov,
       volume = {126},
       number = {5},
        pages = {2268-2280},
          doi = {10.1086/378959},
archivePrefix = {arXiv},
       eprint = {astro-ph/0308357},
 primaryClass = {astro-ph}
}

@article{SAURON_XIV,
    author = {Scott, Nicholas and Cappellari, Michele and Davies, Roger L. and Bacon, R. and De Zeeuw, P. T. and Emsellem, Eric and Falcón-Barroso, Jésus and Krajnović, Davor and Kuntschner, Harald and McDermid, Richard M. and Peletier, Reynier F. and Pipino, Antonio and Sarzi, Marc and Van Den Bosch, Remco C. E. and Van De Ven, Glenn and Van Scherpenzeel, Eveline},
    title = "{The SAURON Project – XIV. No escape from Vesc: a global and local parameter in early-type galaxy evolution}",
    journal = {MNRAS},
    volume = {398},
    number = {4},
    pages = {1835-1857},
    year = {2009},
    month = {09},
    issn = {0035-8711},
    doi = {10.1111/j.1365-2966.2009.15275.x}
}

@ARTICLE{Liang_2024,
       author = {Liang, Fu-Heng and Bureau, Martin and Davis, Timothy},
        title = "{WISDOM Project – ??. Giant molecular clouds in the lenticular galaxy NGC 1387}",
      journal = {\mnras},
         year = 2025,
         volume = "submitted"
}

@article{Lelli_2022,
   title={WISDOM Project – XIII. Feeding molecular gas to the supermassive black hole in the starburst AGN-host galaxy Fairall 49},
   volume={516},
   DOI={10.1093/mnras/stac2493},
   number={3},
   journal={MNRAS},
   publisher={Oxford University Press (OUP)},
   author={Lelli, Federico and Davis, Timothy A and Bureau, Martin and Cappellari, Michele and Liu, Lijie and Ruffa, Ilaria and Smith, Mark D and Williams, Thomas G},
   year={2022},
   month=sep, 
    pages={4066} }

@article{Dominiak_2025,
    author = {Dominiak, Pandora and Cappellari, Michele and Bureau, Martin and Davis, Timothy A and Sarzi, Marc and Ruffa, Ilaria and Iguchi, Satoru and Williams, Thomas G and Zhang, Hengyue},
    title = {WISDOM Project–XXVI. Cross-checking supermassive black hole mass estimates from ALMA CO gas kinematics and SINFONI stellar kinematics in the galaxy NGC 4751},
    journal = {MNRAS},
    volume = {542},
    number = {3},
    pages = {2039-2059},
    year = {2025},
    month = {08},
    doi = {10.1093/mnras/staf1338}
}

@ARTICLE{Dominiak_2024_MASSIVE,
       author = {{Dominiak}, Pandora and {Bureau}, Martin and {Davis}, Timothy A. and {Ma}, Chung-Pei and {Greene}, Jenny E. and {Gu}, Meng},
        title = "{The MASSIVE survey – XIX. Molecular gas measurements of the supermassive black hole masses in the elliptical galaxies NGC 1684 and NGC 0997}",
      journal = {\mnras},
         year = 2024,
         doi = "https://doi.org/10.1093/mnras/stae314"
}

@article{Iodice_2019,
	author = {Iodice, E. and {Spavone, M.} and {Capaccioli, M.} and {Peletier, R. F.} and {van de Ven, G.} and {Napolitano, N. R.} and {Hilker, M.} and {Mieske, S.} and {Smith, R.} and {Pasquali, A.} and {Limatola, L.} and {Grado, A.} and {Venhola, A.} and {Cantiello, M.} and {Paolillo, M.} and {Falcon-Barroso, J.} and {D\'{}Abrusco, R.} and {Schipani, P.}},
	title = {The Fornax Deep Survey with the VST - V. Exploring the faintest regions of the bright early-type galaxies inside the virial radius},
	DOI= "10.1051/0004-6361/201833741",
	journal = {A\&A},
	year = 2019,
	volume = 623,
	pages = "A1",
}

@ARTICLE{Cappellari_2020,
       author = {{Cappellari}, Michele},
        title = "{Efficient solution of the anisotropic spherically aligned axisymmetric Jeans equations of stellar hydrodynamics for galactic dynamics}",
      journal = {\mnras},
         year = 2020,
        month = jun,
       volume = {494},
       number = {4},
        pages = {4819-4837},
          doi = {10.1093/mnras/staa959},
archivePrefix = {arXiv},
       eprint = {1907.09894},
 primaryClass = {astro-ph.GA}
}

@misc{Kabasares_2024, 
      author={Kyle M. Kabasares and Jonathan H. Cohn and Aaron J. Barth and Benjamin D. Boizelle and Jared Davidson and Janelle M. Sy and Jeysen Flores-Velázquez and Silvana C. Delgado Andrade and David A. Buote and Jonelle L. Walsh and Andrew J. Baker and Jeremy Darling and Luis C. Ho},
      year={2024},
      eprint={2403.00181},
      archivePrefix={arXiv},
      primaryClass={astro-ph.GA}
}

@ARTICLE{Jorsater_1995,
       author = {{Jorsater}, Steven and {van Moorsel}, Gustaaf A.},
        title = "{High Resolution Neutral Hydrogen Observations of the Barred Spiral Galaxy NGC 1365}",
      journal = {\aj},
         year = 1995,
        month = nov,
       volume = {110},
        pages = {2037},
          doi = {10.1086/117668}
}

@ARTICLE{Cohn_2024,
       author = {{Cohn}, Jonathan H. and {Curliss}, Maeve and {Walsh}, Jonelle L. and {Kabasares}, Kyle M. and {Boizelle}, Benjamin D. and {Barth}, Aaron J. and {Gebhardt}, Karl and {G{\"u}ltekin}, Kayhan and {Buote}, David A. and {Darling}, Jeremy and {Baker}, Andrew J. and {Ho}, Luis C.},
        title = "{Modeling ALMA Observations of the Warped Molecular Gas Disk in the Red Nugget Relic Galaxy NGC 384}",
      journal = {\apj},
         year = 2024,
        month = nov,
       volume = {975},
       number = {2},
          eid = {179},
        pages = {179},
          doi = {10.3847/1538-4357/ad7bb0},
archivePrefix = {arXiv},
       eprint = {2409.08812},
 primaryClass = {astro-ph.GA}
}

@article{Davis_2014,
    author = {Davis, Timothy A.},
    title = "{A figure of merit for black hole mass measurements with molecular gas}",
    journal = {MNRAS},
    volume = {443},
    number = {1},
    pages = {911-918},
    year = {2014},
    month = {07},
    issn = {0035-8711},
    doi = {10.1093/mnras/stu1163}
}

@InProceedings{Rybicki1987,
  author    = {Rybicki, G. B.},
  booktitle = {Structure and Dynamics of Elliptical Galaxies},
  title     = {Deprojection of Galaxies - how much can BE Learned},
  year      = {1987},
  address   = {Dordrecht},
  editor    = {{de Zeeuw}, P.~T.},
  pages     = {397},
  publisher = {D. Reidel},
  series    = {IAU Symposium},
  volume    = {127},
  doi       = {10.1007/978-94-009-3971-4_41}
}

@Article{vandenBosch2009,
  author        = {van den Bosch, R. C. E. and van de Ven, G.},
  journal       = {\mnras},
  title         = {Recovering the intrinsic shape of early-type galaxies},
  year          = {2009},
  month         = sep,
  pages         = {1117--1128},
  volume        = {398},
  archiveprefix = {arXiv},
  doi           = {10.1111/j.1365-2966.2009.15177.x},
  eprint        = {0811.3474},
}

@article{WISDOM_XVI,
    author = {Elford, Jacob S and Davis, Timothy A and Ruffa, Ilaria and Bureau, Martin and Cappellari, Michele and Gensior, Jindra and Iguchi, Satoru and Liang, Fu-Heng and Liu, Lijie and Lu, Anan and Williams, Thomas G},
    title = {WISDOM Project - XVI. The link between circumnuclear molecular gas reservoirs and active galactic nucleus fuelling},
    journal = {MNRAS},
    volume = {528},
    number = {1},
    pages = {319-336},
    year = {2023},
    month = {12},
    doi = {10.1093/mnras/stad4006}
}

@ARTICLE{Fitzpatrick_1999,
       author = {{Fitzpatrick}, Edward L.},
        title = "{Correcting for the Effects of Interstellar Extinction}",
      journal = {\pasp},
         year = 1999,
        month = jan,
       volume = {111},
       number = {755},
        pages = {63-75},
          doi = {10.1086/316293},
archivePrefix = {arXiv},
       eprint = {astro-ph/9809387},
 primaryClass = {astro-ph}
}

@ARTICLE{Cohn_2023,
       author = {{Cohn}, Jonathan H. and {Curliss}, Maeve and {Walsh}, Jonelle L. and {Kabasares}, Kyle M. and {Boizelle}, Benjamin D. and {Barth}, Aaron J. and {Gebhardt}, Karl and {G{\"u}ltekin}, Kayhan and {Y{\i}ld{\i}r{\i}m}, Ak{\i}n and {Buote}, David A. and {Darling}, Jeremy and {Baker}, Andrew J. and {Ho}, Luis C.},
        title = "{ALMA Gas-dynamical Mass Measurement of the Supermassive Black Hole in the Red Nugget Relic Galaxy PGC 11179}",
      journal = {\apj},
         year = 2023,
        month = dec,
       volume = {958},
       number = {2},
          eid = {186},
        pages = {186},
          doi = {10.3847/1538-4357/ad029d}
}

@article{Harrison_2018,
   title={AGN outflows and feedback twenty years on},
   volume={2},
   ISSN={2397-3366},
   DOI={10.1038/s41550-018-0403-6},
   number={3},
   journal={Nature Astronomy},
   publisher={Springer Science and Business Media LLC},
   author={Harrison, C. M. and Costa, T. and Tadhunter, C. N. and Flütsch, A. and Kakkad, D. and Perna, M. and Vietri, G.},
   year={2018},
   month=feb, pages={198–205} 
}

@ARTICLE{Donofrio_2021,

AUTHOR={D’Onofrio, Mauro  and Marziani, Paola  and Chiosi, Cesare },

TITLE={Past, Present, and Future of the Scaling Relations of Galaxies and Active Galactic Nuclei},

JOURNAL={FSPAS},

VOLUME={8},

YEAR={2021},

DOI={10.3389/fspas.2021.694554},

ISSN={2296-987X}}

@article{Cappellari_2009,
    author = {Cappellari, Michele and Neumayer, N. and Reunanen, J. and Van Der Werf, P. P. and De Zeeuw, P. T. and Rix, H.-W.},
    title = {The mass of the black hole in Centaurus A from SINFONI AO-assisted integral-field observations of stellar kinematics},
    journal = {\mnras},
    volume = {394},
    number = {2},
    pages = {660-674},
    year = {2009},
    month = {03},
    issn = {0035-8711},
    doi = {10.1111/j.1365-2966.2008.14377.x}
}

@article{vanderMarel_1998,
 doi = {10.1086/300593},
 year = {1998},

 volume = {116},
 number = {5},
 pages = {2220},
 author = {Roeland P. van der Marel and Frank C. van den Bosch},
 title = {Evidence for a 3 × 108M☉ Black Hole in NGC 7052 fromHubble Space Telescopebservations of the Nuclear GasDisk*},
 journal = {AJ}
}

@ARTICLE{Ferrarese_1996,
       author = {{Ferrarese}, Laura and {Ford}, Holland C. and {Jaffe}, Walter},
        title = "{Evidence for a Massive Black Hole in the Active Galaxy NGC 4261 from Hubble Space Telescope Images and Spectra}",
      journal = {\apj},
         year = 1996,
        month = oct,
       volume = {470},
        pages = {444},
          doi = {10.1086/177876}
}

@article{Pesce_2020,
doi = {10.3847/2041-8213/ab75f0},
url = {https://dx.doi.org/10.3847/2041-8213/ab75f0},
year = {2020},
month = {feb},
publisher = {The American Astronomical Society},
volume = {891},
number = {1},
pages = {L1},
author = {D. W. Pesce and J. A. Braatz and M. J. Reid and A. G. Riess and D. Scolnic and J. J. Condon and F. Gao and C. Henkel and C. M. V. Impellizzeri and C. Y. Kuo and K. Y. Lo},
title = {The Megamaser Cosmology Project. XIII. Combined Hubble Constant Constraints},
journal = {ApJL}
}

@Article{Ruffa_2024,
AUTHOR = {Ruffa, Ilaria and Davis, Timothy A.},
TITLE = {Molecular Gas Kinematics in Local Early-Type Galaxies with ALMA},
JOURNAL = {Galaxies},
VOLUME = {12},
YEAR = {2024},
NUMBER = {4},
ARTICLE-NUMBER = {36},
ISSN = {2075-4434},
DOI = {10.3390/galaxies12040036}
}

@ARTICLE{McQuinn_2021,
       author = {{McQuinn}, Kristen B.~W. and {Telidevara}, Anjana K. and {Fuson}, Jackson and {Adams}, Elizabeth A.~K. and {Cannon}, John M. and {Skillman}, Evan D. and {Dolphin}, Andrew E. and {Haynes}, Martha P. and {Rhode}, Katherine L. and {Salzer}, John. J. and {Giovanelli}, Riccardo and {Gordon}, Alex J.~R.},
        title = "{Galaxy Properties at the Faint End of the H I Mass Function}",
      journal = {\apj},
         year = 2021,
        month = sep,
       volume = {918},
       number = {1},
          eid = {23},
        pages = {23},
          doi = {10.3847/1538-4357/ac03ae},
archivePrefix = {arXiv},
       eprint = {2105.05100},
 primaryClass = {astro-ph.GA}
}

@ARTICLE{Riess_2024,
       author = {{Riess}, Adam G. and {Scolnic}, Dan and {Anand}, Gagandeep S. and {Breuval}, Louise and {Casertano}, Stefano and {Macri}, Lucas M. and {Li}, Siyang and {Yuan}, Wenlong and {Huang}, Caroline D. and {Jha}, Saurabh and {Murakami}, Yukei S. and {Beaton}, Rachael and {Brout}, Dillon and {Wu}, Tianrui and {Addison}, Graeme E. and {Bennett}, Charles and {Anderson}, Richard I. and {Filippenko}, Alexei V. and {Carr}, Anthony},
        title = "{JWST Validates HST Distance Measurements: Selection of Supernova Subsample Explains Differences in JWST Estimates of Local H0}",
      journal = {arXiv e-prints},
         year = 2024,
        month = aug,
          eid = {arXiv:2408.11770},
        pages = {arXiv:2408.11770},
          doi = {10.48550/arXiv.2408.11770},
archivePrefix = {arXiv},
       eprint = {2408.11770},
 primaryClass = {astro-ph.CO}
}

@misc{Cantiello_2023,
      title={Surface Brightness Fluctuations}, 
      author={Michele Cantiello and John P. Blakeslee},
      year={2023},
      eprint={2307.03116},
      archivePrefix={arXiv},
      primaryClass={astro-ph.CO},
      url={https://arxiv.org/abs/2307.03116}, 
}

@article{Napolitano_2022,
	author = {{Napolitano}, N. R. and {Gatto, M.} and {Spiniello, C.} and {Cantiello, M.} and {Hilker, M.} and {Arnaboldi, M.} and {Tortora, C.} and {Chaturvedi, A.} and {D’Abrusco, R.} and {Li, R.} and {Paolillo, M.} and {Peletier, R.} and {Saifollahi, T.} and {Spavone, M.} and {Venhola, A.} and {Capaccioli, M.} and {Longo, G.}},
	title = {The Fornax Cluster VLT Spectroscopic Survey - IV. Cold kinematical substructures in the Fornax core from COSTA},
	DOI= "10.1051/0004-6361/202141872",
	url= "https://doi.org/10.1051/0004-6361/202141872",
	journal = {A&A},
	year = 2022,
	volume = 657,
	pages = "A94",
}

@article{WISDOM_XXII,
    author = {Zhang, Hengyue and Bureau, Martin and Ruffa, Ilaria and Cappellari, Michele and Davis, Timothy A and Dominiak, Pandora and Elford, Jacob S and Iguchi, Satoru and Lelli, Federico and Sarzi, Marc and Williams, Thomas G},
    title = {WISDOM Project – XXII. A 5 per cent precision CO-dynamical supermassive black hole mass measurement in the galaxy NGC 383},
    journal = {MNRAS},
    volume = {537},
    number = {1},
    pages = {520-536},
    year = {2025},
    month = {01},
    doi = {10.1093/mnras/staf055}
}

@article{WISDOM_XXV,
    author = {Zhang, Hengyue and Bureau, Martin and Ruffa, Ilaria and Davis, Timothy A and Dominiak, Pandora and Elford, Jacob S and Lelli, Federico and Williams, Thomas G},
    title = {WISDOM Project – XXV. Improving the CO-dynamical supermassive black hole mass measurement in the galaxy NGC 1574 using high spatial resolution ALMA observations},
    journal = {MNRAS},
    pages = {staf1161},
    year = {2025},
    month = {07},
    doi = {10.1093/mnras/staf1161}
}

@ARTICLE{Andrae_2010,
       author = {{Andrae}, Rene and {Schulze-Hartung}, Tim and {Melchior}, Peter},
        title = "{Dos and don'ts of reduced chi-squared}",
      journal = {arXiv e-prints},
         year = 2010,
        month = dec,
          eid = {arXiv:1012.3754},
        pages = {arXiv:1012.3754},
          doi = {10.48550/arXiv.1012.3754}
}

%%%%%%%%%%%%%%%%%%%%%%%%%%%%%%%%%%%%%%%%%%%%%%%%%%%%%%%%%%%%%%%%%%%%%%%%%%%%%%%%%%%%%%%%%%%%%%%%%%%%

%%%%%%%%%%%%%%%%%%%%%%%%%%%%%%%%%%%%%%%%%%%%%%%%%%%%%%%%%%%%%%%%%%%%%%%%%%%%%%%%%%%%%%%%%%%%%%%%%%%%

% Don't change these lines
\bsp	% typesetting comment
\label{lastpage}

%\begin{thebibliography}{99}
%\end{thebibliography}

\end{document}